\documentclass[useAMS,usenatbib]{mnras}
\pdfoutput=1
\usepackage{graphicx}
\usepackage{amsmath}
\usepackage{amssymb}
\usepackage{comment}
\usepackage{subfigure}

\newcommand{\aap}{A\&A}
\newcommand{\araa}{ARA\&A}
\newcommand{\apj}{ApJ}
\newcommand{\apjl}{ApJ}

\newcommand{\mnras}{MNRAS}

\newcommand{\msun}{\ensuremath{{\rm M}_{\sun}}}
\newcommand{\apjs}{ApJS}
\newcommand{\physrep}{Phys. Rep.}

\newcommand{\na}{New Astron.}
\newcommand{\prd}{PRD}

\newcommand{\pasp}{PASP} 
\newcommand{\sovast}{Soviet Astronomy}

\include{defns}
\def\gsim{\;\rlap{\lower 2.5pt
 \hbox{$\sim$}}\raise 1.5pt\hbox{$>$}\;}
\def\lsim{\;\rlap{\lower 2.5pt
   \hbox{$\sim$}}\raise 1.5pt\hbox{$<$}\;}
\defcitealias{liu09a}{Liu et al. (2009a)}
\defcitealias{liu09b}{Liu et al (2009b)}

\newcommand{\rta}{r_{\rm ta}}
\newcommand{\rshock}{r_{\rm s}}

\title[On the Apparent Power Law in CDM Halo PPSD Profiles]{On the Apparent Power Law in CDM Halo Pseudo Phase Space Density Profiles}
\author[Nadler, Oh \& Ji]{Ethan O.\ Nadler, S.\ Peng Oh, Suoqing Ji \\
Dept. of Physics, University of California, Santa Barbara, CA 93106, USA.}

\begin{document}
\bibliographystyle{mnras}

\pagerange{000--000} \pubyear{0000}
\maketitle

\label{firstpage}

\begin{abstract}
We investigate the apparent power-law scaling of the pseudo phase space density (PPSD) in CDM halos. We study fluid collapse, using the close analogy between the gas entropy and the PPSD in the fluid approximation. Our hydrodynamic calculations allow for a precise evaluation of logarithmic derivatives. For scale-free initial conditions, entropy is a power law in Lagrangian (mass) coordinates, but {\it not} in Eulerian (radial) coordinates. The deviation from a radial power law arises from incomplete hydrostatic equilibrium (HSE), linked to bulk inflow and mass accretion, and the convergence to the asymptotic central power-law slope is very slow. For more realistic collapse, entropy is not a power law with either radius or mass due to deviations from HSE and scale-dependent initial conditions. Instead, it is a slowly rolling power law that appears approximately linear on a log-log plot. Our fluid calculations recover PPSD power-law slopes and residual amplitudes similar to N-body simulations, indicating that deviations from a power law are not numerical artefacts. In addition, we find that realistic collapse is not self-similar: scale lengths such as the shock radius and the turnaround radius are not power-law functions of time. We therefore argue that the apparent power-law PPSD cannot be used to make detailed dynamical inferences or extrapolate halo profiles inward, and that it does not indicate any hidden integrals of motion. We also suggest that the apparent agreement between the PPSD and the asymptotic Bertschinger slope is purely coincidental.
\end{abstract}



\section{Introduction}
\label{section:intro}

What determines the final state of a large collection of particles interacting under Newtonian gravity? Despite decades of effort, a satisfactory answer to this question remains elusive. The difficulty is due primarily to the long range, unshielded nature of gravity. While gaseous systems with short-range interactions quickly relax to a Maxwellian equilibrium, self-gravitating systems have a collisional relaxation time that scales with the number of particles, making two-body relaxation essentially negligible on galactic scales. Instead, collective processes such as phase-mixing and violent relaxation govern the approach to equilibrium. The fact that relaxation is incomplete is clear because the results of N-body simulations depend on initial conditions, such as the initial virial ratio \citep{van-albada82}. Indeed, the only distribution function consistent with complete relaxation results in a singular isothermal profile with infinite mass \citep{lynden-bell67}. The entire notion of thermodynamic equilibrium in a self-gravitating system is suspect, since at fixed mass and energy one can always arbitrarily increase the entropy of a system by increasing its central concentration \citep{lynden-bell68,tremaine86}. These problems arise directly from the fact that gravitational systems are non-extensive and non-additive\footnote{The sum of the energies of individual components is not equal to the energy of the entire system.}. Similar difficulties appear in other systems with long-range interactions for which standard Boltzmann-Gibbs statistics do not apply; examples include 2D hydrodynamic systems, 2D elastic systems, and charged or dipolar systems \citep{campa09}\footnote{Efforts to apply non-extensive statistics (e.g., \citealt{tsallis01}) have not been particularly successful.}.

Despite these theoretical difficulties, there is now overwhelming evidence from N-body simulations that structure formation results in nearly universal CDM halo profiles (see \citet{frenk12} for a review); the most famous of these features is the density profile, as described by either the NFW \citep{navarro97} or Einasto \citep{navarro04} fitting formulae. The overall shape of the density profile is independent of the initial fluctuation spectrum, of halo mass and formation epoch, and of cosmological parameters. Indeed, similar profiles even arise in the absence of hierarchical growth \citep{huss99,moore99,wang09}. The logarithmic slope of the density profile continually steepens from $\sim -1$ or even shallower in the innermost regions to $\sim -3$ at the halo outskirts. These universal density profiles in turn imply universal circular velocity profiles. In the original NFW fit, the only weak scale dependence is encapsulated by the halo concentration $c=r_{\rm 200}/r_{-2}$ (where $r_{-2}$ is the radius at which $\text{d}(\text{log}\rho)/\text{d}(\text{log}r)=-2$), which has been shown to reflect halo formation time.

Most work to date has focused on explaining the origin of the dark matter density profile, which for instance has been attributed to tidal disruption of substructure \citep{syer98,subramanian00,dekel03a,dekel03b}, the shape of the matter power spectrum \citep{nusser01}, a modulation of the accretion rate \citep{lu06}, the adiabatic contraction of the peaks of Gaussian random fields \citep{dalal10}, and conservation of orbital actions \citep{pontzen13}. At this point, a consensus explanation seems unlikely \citep{frenk12}. However, there are other intriguing regularities that provide insights into halo structure. Halos clearly show common trends in orbital anisotropy $\beta(r) = 1 - \sigma^{2}_{\theta}/\sigma^{2}_{r}$: they are quasi-isotropic ($\beta \approx 0$) at the centre and radially anisotropic outwards, with $\beta \approx 0.25$ at $r_{-2}$ and $\beta \approx 0.5$ farther out \citep{navarro10,ludlow10}. Within $r_{-2}$, the anisotropy parameter $\beta(r)$ is correlated with the logarithmic slope of the density profile $\gamma(r)$ via $\beta(r) \approx -0.15 - 0.2 \gamma(r)$ \citep{hansen06-DM}. Similarly, the cumulative distribution of specific angular momentum $j$ can be fit by a universal function $M(<j) \propto j$ (with a flattening at large $j$), albeit with large scatter \citep{bullock01,bett10}. 

Most remarkably, the spherically averaged quantity $Q(r) \equiv \rho/\sigma^{3}$ follows an approximate power law $Q(r) \propto r^{-\alpha}$, where $\alpha \approx 1.9$, over three orders of magnitude in radius over the entire relaxed halo profile \citep{taylor01,navarro10,ludlow10}. This relationship holds despite the fact that neither $\rho(r)$ nor $\sigma(r)$ are power laws; in fact, these quantities vary substantially from halo to halo. Like the NFW profile, $Q(r)$ is robust to mergers and whether structure formation is hierarchical \citep{hoffman07,wang09}, and its slope only depends very weakly on the power spectrum \citep{knollmann08}, so its origin is likely unrelated to any of these. As $r \rightarrow 0$, $Q(r)$ asymptotically approaches a similar power law to the one predicted by the 1D model of self-similar collapse in \citet{bertschinger85}, which otherwise disagrees with simulations since it predicts incorrect asymptotic power-law slopes for the density and velocity dispersion profiles. While $Q(r)$ has units of phase space density, it is in fact a ratio of moments of the distribution function\footnote{Indeed, the coarse-grained distribution function $\bar{f}$, which is accessible in simulations, is not a power law and flattens considerably at large radii \citep{vass09}.}, and hence is often dubbed the `pseudo phase space density'. A very similar power law holds if one substitutes the radial velocity dispersion for the total velocity dispersion \citep{dehnen05,navarro10}.

The origin of this remarkable power-law relationship is not known, and understanding it could pay considerable dividends. In contrast to empirical fitting formulae such as the NFW and Einasto profiles, a scale-free power law emerges naturally in a similarity solution, generally as the result of conserved quantities. A true power-law $Q(r)$ would therefore be a `fundamental' universal feature of CDM halos, in the sense that its power-law index should be calculable from the equations and boundary conditions governing halo collapse. If this is the case, the relationship could be used when numerical resolution peters out. For instance, density profiles must differ from a simple Einasto form if a power-law $Q(r)$ holds at all radii \citep{ma09}, but current numerical simulations cannot test this claim \citep{ludlow11}. In addition, many of the present universal features of equilibrium CDM halos would follow as a consequence of a power-law $Q(r)$. In particular, the Jeans equation, which essentially describes hydrostatic equilibrium, 
\begin{equation} 
\frac{1}{\rho}\frac{\text{d}}{\text{d}\,r}\left( \rho \sigma_{r}^{2} \right) + 2 \beta \frac{\sigma_{r}^{2}}{r} + \frac{G M(<r)}{r^{2}}=0,\label{eqn:1.1}\end{equation}
has three free functions: $\rho(r)$, $\sigma^{2}_{r}(r)$, and $\beta(r)$. $Q(r)$ relates $\rho(r)$ and $\sigma_{r}^{2}(r)$ via an effective `equation of state', so only $\beta(r)$ needs to be specified to close the system of equations. Assuming orbital isotropy ($\beta(r) \approx 0$), as appropriate for halo centres, the observed power-law $Q(r) \propto r^{-\alpha}$ with $\alpha\approx 1.9$ is the only solution of equation \ref{eqn:1.1} that yields a physical (non-negative and non-truncated) density profile, and this profile agrees with the NFW fit \citep{taylor01,hansen04-dm,dehnen05}. Additional assumptions are needed for the more general anisotropic case, but if one assumes $\beta(r) \propto \gamma(r)$, as seen in numerical simulations, then one can recover density, velocity dispersion, and anisotropy profiles  in good agreement with cosmological simulations, {\it independent} of the assumed slope of the $\beta(r)-\gamma(r)$ relation \citep{dehnen05}.

A power law pseudo phase space density also has some observational support. From member galaxy kinematics and from lensing, X-ray, and kinematical mass profiles, a power-law $Q(r)$ in agreement with simulations has been inferred for galaxy cluster Abell 2142 \citep{munari14}. Measurements of the mass density $\rho(r)$ and the radial stellar velocity dispersion $\sigma_{r,*}(r)$ of $\sim 2000$ SDSS elliptical galaxies yield a power-law $Q(r) \propto r^{-\chi}$, with $\chi = 1.860 \pm 0.035$, in remarkable agreement with values for simulated CDM halos \citep{chae14,chae14a}. This is particularly striking because density profiles for ellipticals are not universal and differ sharply from halo density profiles, largely because the stellar components of ellipticals are influenced by dissipative processes (e.g., gas cooling to form discs or the merger of disc galaxies to form a bulge). The fact that $Q(r)$ follows a power law with a similar slope for elliptical galaxies, galaxy clusters, and dark matter halos reinforces the view that $Q(r)$ is a more universal and fundamental feature of gravitational collapse than $\rho(r)$.

The apparent power-law nature of $Q(r)$ can therefore deliver powerful insights into the universality of CDM halo structure. However, a distinction must be drawn between approximate and exact power laws. Exact power laws often have deep physical significance; the associated scale invariance generally arises from conserved integrals of motion (e.g., energy conservation in Taylor-Sedov explosions). In the case of `Type I similarity', the (rational) power-law exponent follows from simple dimensional analysis. In `Type II similarity', the integrals of motion and the (irrational) power-law exponent can only be obtained by solving an eigenvalue problem; this type of self-similarity arises from invariance under the renormalisation group \citep{goldenfeld92,barenblatt96}. On the other hand, the power-law nature of $Q(r)$ has been inferred simply by fitting noisy data on log-log plots, and it is notoriously easy to obtain spurious power-law relationships in this manner (e.g., see \citet{clauset09} for the plethora of pitfalls associated with inferring power-law distributions). State of the art simulations show $\sim 20\%$ deviations from the best-fitting power-law $Q(r)$, which moreover varies systematically with radius. Given the current numerical limitations, it is difficult to assess the significance of these deviations. Thus, simple toy models that can be solved to high precision (and benchmarked by comparison to exact self-similar calculations) are valuable. In this paper, we show that simple 1D calculations of fluid collapse satisfy these criteria. While these models do not possess all of the complex degrees of freedom encompassed by cosmological N-body simulations, they are very illustrative. We show that the entropy profiles from fluid calculations are remarkably similar to the pseudo phase space density profiles of CDM halos in N-body simulations; in particular, they have similar straight-line fits on log-log plots. However, these profiles deviate from true power laws for simple physical reasons, even for idealised collapse with scale-free initial conditions. We therefore argue that the power-law fits to $Q(r)$ do not have deep physical significance.

This paper is organised as follows. In \S\ref{sect:fluid}, we discuss the applicability of the fluid approximation to CDM halos. In \S\ref{sect:methods}, we describe our method for simulating realistic halo collapse and extracting pseudo phase space density profiles via the fluid approximation. We also consider halo collapse with scale-free initial conditions and compare our simulation with an analytic calculation. We present our results in \S\ref{sect:results} and conclude in \S\ref{sect:conclusions}.


\section{The Fluid Approximation} 
\label{sect:fluid}

In this paper, we use the `fluid approximation' for dark matter. This technique utilises the fact that the zeroth, first, and second moments of the collisionless Boltzmann equation resemble fluid equations for the conservation of mass, momentum, and energy. The main difference with a true fluid that has scalar pressure $p$ is that collisionless systems are supported against gravity by gradients in the (generally anisotropic) stress tensor $\rho \sigma_{ij}^{2}$. The fluid approximation has been used in stellar dynamics to study the gravothermal catastrophe \citep{larson70,lynden-bell80}, giving results in good agreement with numerical Fokker-Planck and N-body calculations. More recently, it has been used to study the collapse of collisionless \citep{teyssier97,subramanian00,lapi11} and self-interacting \citep{ahn05} CDM halos. Given the same assumptions as the classic particle-based similarity solutions of \citet{fillmore84} and \citet{bertschinger85} (1D radial infall and scale-free initial conditions in an Einstein de-Sitter universe), fluid calculations correctly recover the same density and velocity profiles. 

The solution procedure is as follows. Initially, the cold fluid obeys the standard parametric relations for turnaround and free-fall onto an overdensity \citep{peebles80}. The hyperbolic fluid equations undergo a discontinuous `shock' at the radius where shell-crossing would occur, and the single-stream flow becomes multi-stream. This can be treated by the standard Rankine-Hugoniot shock jump conditions. In the particle-based approach, this corresponds to the first caustic, and relaxation (i.e., phase-mixing) occurs in a thicker region, rather than instantaneously, due to successive shell-crossings. The shock thermalizes the fluid, converting ordered bulk motion into random motions. 

The assumption that the infinite BBGKY hierarchy can be truncated at its second moment deserves justification. It amounts to the assumption that $v_{\rm r}$ has a symmetric distribution, and in particular that it is skewless\footnote{By spherical symmetry, $v_{\theta}$ and $v_{\phi}$ are already skewless.}: $\langle (v_{r}-\bar{v}_{r})^{3} \rangle = 0$. An asymmetric distribution implies net radial energy flow through the system, and in general skewness is zero except near caustics. In the particle approach, this can be seen directly from the symmetric distribution of radial velocities in phase diagrams. The hypothesis that the radial action is an adiabatic invariant \citep{fillmore84} is equivalent to our assumption that the fluid does not experience shell-crossing, since in both cases the enclosed mass within a shell is constant on orbital timescales \citep{teyssier97}. 

For our purposes, the fluid approximation has a number of advantages. Self-similar calculations are straightforward in the fluid approximation; agreement with or departure from collisionless N-body results can shed further insights. Indeed, the case of pure fluid collapse is an interesting problem in its own right and can be tested against hydrodynamic simulations. Furthermore, since we focus on the pseudo phase space density $Q(r)$, the fluid approximation allows us to exploit a very interesting and useful analogy with the gas entropy $k_{\rm g} = 3 k_{\rm B} T_{\rm g} \rho_{g}^{-2/3} (\mu m_{\rm p})^{-1} = \sigma_{\rm g}^{2} \rho_{\rm g}^{-2/3}$. Simulations of galaxy clusters have shown that, outside the central core (where gas entropy is often flat\footnote{The cluster core is often affected by non-gravitational processes such as AGN feedback and radiative cooling. Even in adiabatic simulations, results for the core entropy depend on the chosen numerical method \citep{mitchell09}, likely due to different treatments for processes such as turbulent mixing. In SPH simulations, where mixing -- which is absent in a collisionless fluid -- is suppressed, the entropy profile remains a power law.}), the analogous quantity $k_{\rm DM} = \sigma_{\rm DM}^{2} \rho_{\rm DM}^{-2/3}$ matches the gas entropy in both its power-law radial scaling ($k_{\rm DM} \propto k_{\rm g} \propto r^{1.2}$) and in its normalisation, once gas bulk motions are taken into account \citep{faltenbacher07}. A similar power-law entropy profile $k_{\rm g} \propto r^{1.21 \pm 0.39}$ is inferred from X-ray observations \citep{cavagnolo09}. A quick heuristic way to understand the power-law gas entropy profile follows from the accretion history of the cluster \citep{tozzi01,voit03}. The gas entropy is a power-law function of mass, $k_{g}(M_{\rm g}) \propto M_{\rm g}^{\gamma}$, since $k_{g} \propto T/\rho^{2/3} \propto v_{\rm in}^{2}/\rho^{2/3} \propto ({GM/R})/\rho^{2/3} \propto M^{2/3} (1+z)^{3} \propto (M t)^{2/3}  \propto M^{1.0-1.4}$, where $v_{\rm in}$ is the infall velocity at the shock and $M \propto t^{\eta}$ with $0.9 \le \eta \le 1.9$ for a $\Omega_{\rm m}=0.3$ flat cosmology with the relevant power spectrum indices $-2 \le n \le -1$ \citep{voit03}. This estimate agrees with our more detailed calculations (for example, see Figure \ref{fig:2.8}). A running power-law scaling of $k_{g}$ with radius then follows by demanding hydrostatic equilibrium. Using the fluid analogy -- that dark matter entropy is generated at `shocks' (i.e., caustics) via phase-mixing and is thereafter conserved -- we can construct pseudo phase space density profiles for dark matter halos and compare against simulations using $Q(r) \propto k_{\text{DM}}^{-3/2} \propto k_{g}^{-3/2}$. 

An important caveat is in order. The fluid model does not give a clear-cut prescription for the velocity anisotropy $\beta(r)$; in particular, it does not specify the directionality of the random motions generated at the shock (of course, in a true fluid, velocities are isotropic). Tangential velocity dispersions arise from asphericity, tidal torques, and the radial orbit instability (ROI). The ROI is particularly interesting: it arises because purely radial orbits are unstable to precessional motion, leading to a quadrupolar bar-like instability. This in turn generates tangential velocity dispersions that are ultimately stabilising \citep{antonov73, polyachenko81, merritt85, palmer87}. The ROI clearly plays an important role in shaping halo density profiles; in simulations of monolithic collapse where only radial forces are evolved, the density profile is close to isothermal. However, once the ROI is allowed to operate, orbits near the centre are isotropised, the central cusp softens, and density profiles transform into an NFW form \citep{huss99,macmillan06}. Unfortunately, there is currently no successful analytic theory for the ROI (see \citet{merritt99, binney08} for reviews) or how it saturates \citep{adams07}. To incorporate these effects, semi-analytic models have inserted tangential velocity dispersions and/or angular momentum by hand, generally near turnaround \citep{nusser01,barnes05,lu06,zukin10}; suitably tuned, this method reproduces the NFW profile. Fortunately, for our goal of a deeper understanding of the power-law nature of $Q(r)$, a detailed understanding of the ROI is unnecessary. A power-law $Q(r)$ is remarkably robust and arises (both in $\rho/\sigma^{3}$ and $\rho/\sigma_{r}^{3}$) in purely radial collapse \citep{bertschinger85}, in isotropic fluid collapse, 
and in N-body simulations where the ROI develops naturally \citep{macmillan06} or where tangential velocities are manipulated by hand \citep{barnes05,lapi11}. It is therefore unlikely to be sensitive to the details of the ROI.


Even if we want to understand the density profile, the detailed shape of $\beta(r)$ may not be necessary. From a phase-plane analysis of the Jeans equation and its derivatives, \citet{dehnen05} have shown that, if we assume a power-law $Q(r)$ and that $\beta(r) \propto \gamma(r) \equiv {\rm d \, log} \rho/{\rm d \, log} r$ (as seen in simulations; \citet{hansen06-DM}), then we only need to know $\beta_{0}=\beta(r=0)$ to construct an analytic density profile in excellent agreement with simulations. This result is {\it independent} of the assumed slope of the $\beta-\gamma$ relation. 
The velocity distribution is close to isotropic near halo centres ($\beta_{0} \approx 0$, with some scatter) in simulations; this can be understood as the effect of efficient violent relaxation during the rapid collapse phase \citep{lu06}, and it is related to the slope of the initial density perturbation \citep{vogelsberger11}. Also, for the purposes of understanding the central cusp (and extrapolating inward beyond the limits of numerical resolution), the details of $\beta(r)$ at large radii are irrelevant.

In this paper, we only consider spherical symmetry and monolithic collapse. Although adding triaxiality does affect the density and anisotropy profiles \citep{lithwick11}, $Q(r)$ is virtually unaffected, remaining a single power law with an index that depends very weakly on the similarity index of the initial conditions \citep{vogelsberger11}. This perhaps hints that entropy stratification is governed primarily by the gravitational potential, which has greater spherical symmetry than the density profile. The hierarchical nature of structure formation does not appear to be important in determining the structure of CDM halos. As noted above, key features (e.g., NFW-like density profiles and power-law $Q(r)$ profiles) already arise in monolithic collapse simulations. Furthermore, simulations of equal mass merger events -- the most violent possible -- show that mixing is remarkably inefficient and that the overall shapes of the progenitor $Q(r)$ profiles (and the coarse-grained phase space density $\bar{f}(x,v)$) are remarkably well-preserved \citep{vass09}, regardless of the merger geometry or the number of mergers. 


\section{Methods} 
\label{sect:methods}

\subsection{Self-similar Fluid Collapse} 
\label{sect:self-similar} 

In an Einstein-de Sitter (EdS) universe with scale-free initial conditions, the evolution of density perturbations is self-similar. As we demonstrate below, the fluid equations then reduce from a set of PDEs to a set of ODEs that can easily be solved to very high precision. This is useful for two reasons: (i) We can test the reliability of our hydrodynamic code for initial conditions where the exact solution is known; (ii) We can study the entropy profile $k_{\rm g}(r)$ for self-similar collapse. If the entropy profile is a power law, then we can study how it changes if exact self-similarity is broken; if it is \emph{not} a power law, then $k_{\rm g}(r)$ is unlikely to be a power law in the more realistic case! Our treatment closely follows that of \citet{bertschinger85}, who considered radial self-similar fluid collapse. The main new element in our analysis is the focus on the entropy profile $k_{g}(r)$.

As in \citet{fillmore84}, scale-free initial density perturbations can be characterised by
\begin{equation}
\frac{\delta M_{i}}{M_{i}} = \left( \frac{M_{i}}{M_{0}} \right)^{-\epsilon},
\label{eqn:1.2}\end{equation}
where $M_{i}$ is the unperturbed mass, $M_{0}$ is a reference mass, and $\epsilon > 0$ is a constant. The positive mass excess implies that each radial shell is bound and collapses after reaching a maximum turnaround radius $r_{\rm ta}$, which scales with time according to
\begin{equation}
r_{\rm ta} \propto t^{\alpha}; \ \alpha=\frac{2}{3} \left(1+ \frac{1}{3\epsilon} \right). 
\label{eqn:1.3}\end{equation}
Here and throughout this paper, $\alpha$ denotes the power-law index of characteristic scale radii as functions of time (note that $\alpha$ is constant for self-similar collapse). We can derive equation \ref{eqn:1.3} by dimensional analysis as follows. First, note that the initial conditions of equation \ref{eqn:1.2} imply that the specific energy $\delta E \propto G\delta M/r \propto M^{1-\epsilon}/M^{1/3} \propto M^{2/3-\epsilon}$, where $E = \dot{r}^{2}/2-GM/r$ is the energy integral of the equation of motion for each shell. The maximum radius each shell reaches satisfies $\delta E = GM/r_{\rm ta}$, so $r_{\rm ta} \propto M/M^{2/3-\epsilon} = M^{1/3+\epsilon}$. The only characteristic time scale is the free-fall time $t_{\rm ff} \propto \sqrt{1/G\rho} \propto \sqrt{r^{3}/M}$, so we also have $r_{\rm ta} \propto t^{2/3}M^{1/3}$. Thus $M \propto t^{2/3\epsilon}$ and $r_{\rm ta} \propto t^{(2/3)(1+1/3\epsilon)}$. The shock radius of the infalling matter scales identically by self-similarity. 

Since there are no other characteristic scales in the problem, the behaviour of the system is self-similar once rescaled to the turnaround radius or the shock radius. In analogy with \citet{bertschinger85}, we non-dimensionalize the problem by factoring out the time-dependent background density $\bar{\rho}(t)$ and a characteristic scale radius $\rshock(t)$: 
\begin{eqnarray}
\label{eq:fluid_vars1}
\lambda(r,t) &=& \frac{r}{\rshock} \\
v(r,t)&=&\frac{\rshock}{t}V(\lambda) \\
\rho(r,t)&=& \bar{\rho} D(\lambda) \\ 
p(r,t) &=& \bar{\rho} \left(\frac{\rshock}{t}\right)^{2} P(\lambda) \\ 
m(r,t) &=& \frac{4 \pi}{3} \bar{\rho} \rshock^{3} M(\lambda).
\label{eq:fluid_vars5}
\end{eqnarray} 
The usual choice for the scale radius is the turnaround radius $r_{\rm ta}(t)$, but in this paper we will use the shock radius for $\rshock(t)$. Thus, $\lambda$ refers to the radius normalised to the radius of the fluid shock. Note that the ratio $\rshock/r_{\rm ta}$ is constant for self-similar collapse, but this is no longer true for more realistic initial conditions (\S\ref{sect:realistic}). Using equations \ref{eq:fluid_vars1}-\ref{eq:fluid_vars5}, the fluid equations in spherical symmetry become:
\begin{eqnarray} 
\label{eq:fluid_self_similar1} 
(V-\alpha \lambda) D^{\prime} + DV^{\prime} + \frac{2 DV}{\lambda} - 2D = 0 \\
(V-\alpha \lambda) V^{\prime} - (1-\alpha)V = -\frac{P^{\prime}}{D} - \frac{2}{9}\frac{M}{\lambda^{2}} \\
\left(\gamma\frac{D^{\prime}}{D} - \frac{P^{\prime}}{P} \right) \left(V-\alpha \lambda \right) = 2 (\alpha-2+\gamma) \label{eq:fluid_self_similar3} \\
M^{\prime} = 3 \lambda^{2} D, 
\label{eq:fluid_self_similar4} 
\end{eqnarray}
where the non-dimensionalized velocity, density, pressure, and mass profiles are functions of $\lambda$ and primes denote total derivatives with respect to $\lambda.$ Note that $M(\lambda)$ is the mass interior to a fluid element at radius $\lambda.$

The above system of ODEs can be solved numerically. We describe the procedure using the conventional turnaround radius non-dimensionalization. We assume that at the initial time $t_{i}$ the radial velocity field is pure unperturbed Hubble flow. The cold pre-shock fluid is pressureless ($P=0$), so there are only 3 variables: $D$, $V$, and $M$. The initial conditions for these profiles at turnaround are $V=0$, $M=(3\pi/4)^{2}$, and $D=M/(3\epsilon +1)$ \citep{fillmore84}. The equations can be integrated inward from the turnaround radius to the non-dimensionalized shock radius $\lambda_{\rm s} = \rshock/\rta$, whereupon the shock jump conditions
\begin{eqnarray}
\label{eq:shock}
\frac{D_{2}}{D_{1}} &=& \frac{\gamma+1}{\gamma-1} \\ 
\frac{V_{2}-\alpha \lambda}{V_{1}-\alpha \lambda} &=& \frac{D_{1}}{D_{2}} = \frac{\gamma-1}{\gamma+1} \\ 
M_{2} &=& M_{1} \\
P_{2} &=& \frac{2}{\gamma+1} D_{1} (V_{1}- \alpha \lambda_{\rm s})^{2}
\end{eqnarray}
apply,  and the integration can continue inward. Note that $\alpha \lambda_{\rm s}$ is the non-dimensionalized shock speed. The unknown shock eigenvalue $\lambda_{\rm s}$ is varied until the boundary conditions
\begin{equation}
V(0)=0, \ \ \ M(0) = 0,
\end{equation}
are satisfied. $M(0)=0$ guarantees that the mass profile is non-singular -- i.e., there is no central point mass (such as a black hole). $V(0)=0$ immediately follows from this requirement, since a nonzero $V(0)$ in a collisional fluid with no shell-crossing leads to the development of a mass singularity.

Throughout this paper, we use an adiabatic index $\gamma=5/3$. The fact that fluid calculations with $\gamma=5/3$ match calculations of collisionless infall was first noted by \citet{bertschinger85}. He remarked that one might expect a $\gamma=3$ fluid calculation, corresponding to a gas with only one translational degree of freedom, to agree with the 1D collisionless calculation for purely radial infall. This is not the case, since the fluid equations as written above assume an isotropic 3D pressure and use a 3D mass density. Some authors \citep{subramanian00,lapi11} have subsequently formulated the self-similar calculation in 1D with $\gamma=3$ (i.e., using a radial pressure and density), but we use the conventional formulation. 

Equations \ref{eq:fluid_self_similar1}-\ref{eq:fluid_self_similar4} give the mass and entropy integrals \citep{bertschinger85}
\begin{eqnarray}
\label{eq:fluid_integrals1}
M &=& \frac{3}{2-3\alpha} D (V-\alpha \lambda) \lambda^{2} \\
P D^{-\gamma} M^{\zeta} = k_{g}M^{\zeta} &=& {\rm const}, \  \ \zeta = \frac{2 (\alpha + \gamma -2)}{2-3\alpha}
\label{eq:fluid_integrals2}
\end{eqnarray}
where we have defined the fluid entropy $k_{g} = PD^{-\gamma}$. The mass and entropy integrals should be conserved exactly, so they are useful for checking the fidelity of our numerical results.

\subsection{Fluid Collapse Simulation} 
\label{sect:hydro}

To simulate dark matter halo collapse, we use a 1D spherically symmetric Lagrangian hydrodynamics/gravity code, similar to \citet{thoul95}. We initialise concentric fluid shells with a density profile $\rho_{i}(r)$ and a radial velocity profile $v_{i}(r)$ corresponding to pure unperturbed Hubble flow at a fiducial initial redshift $z_{i}$. The self-gravitating system then evolves according to the fluid equations; at the radius where shell-crossing would occur, the fluid undergoes a discontinuous shock, which we treat with an artificial viscosity technique as in \citet{thoul95}. We allow the system to evolve to a final redshift $z_{f}$, which we take as $z_{f} = 0$ throughout. We choose $M_{0} = 10^{12}\ \msun$ for our reference mass.

\begin{figure}
\includegraphics[width=1.0\columnwidth]{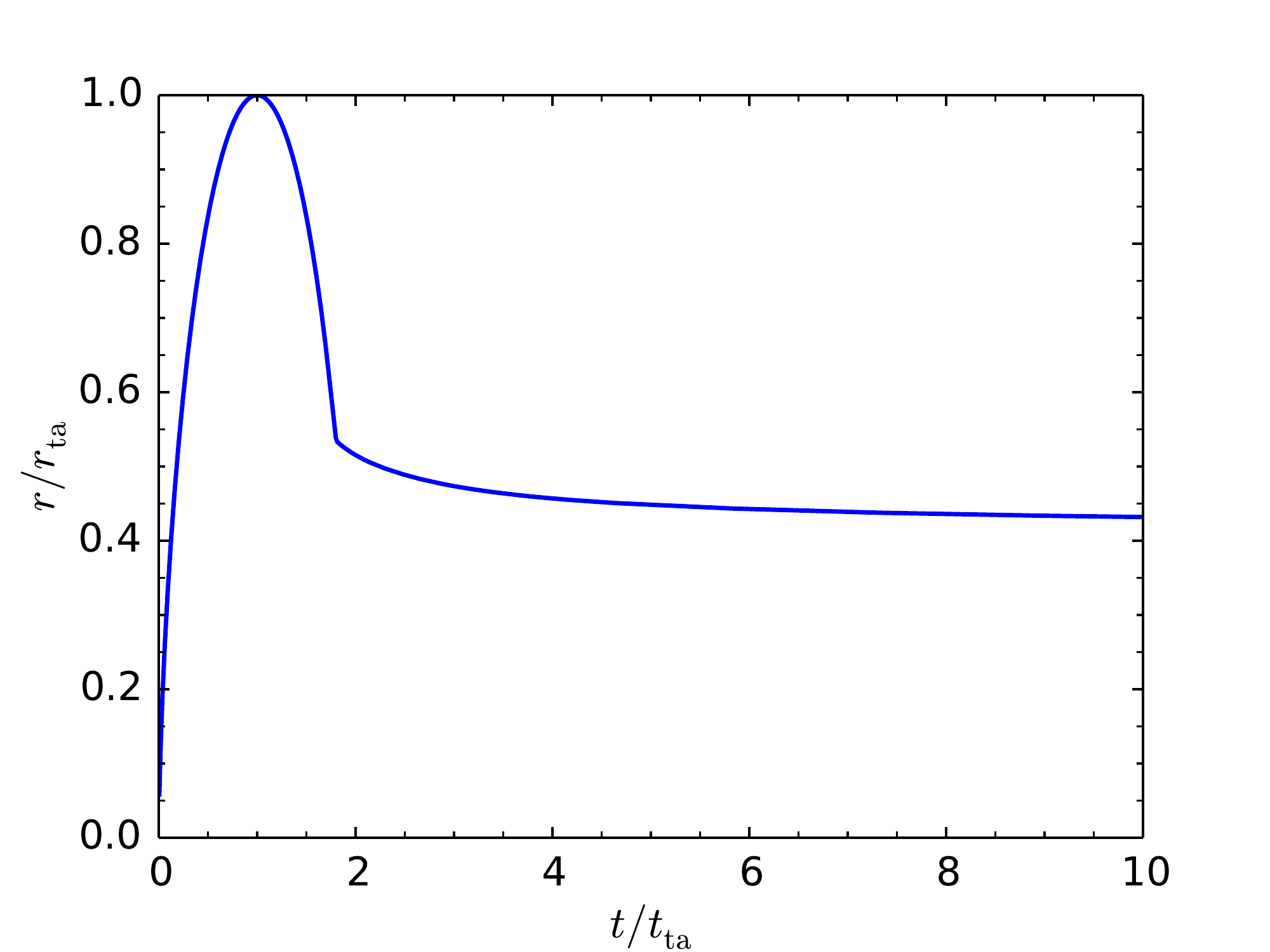}\vspace{2mm}
\includegraphics[width=1.0\columnwidth]{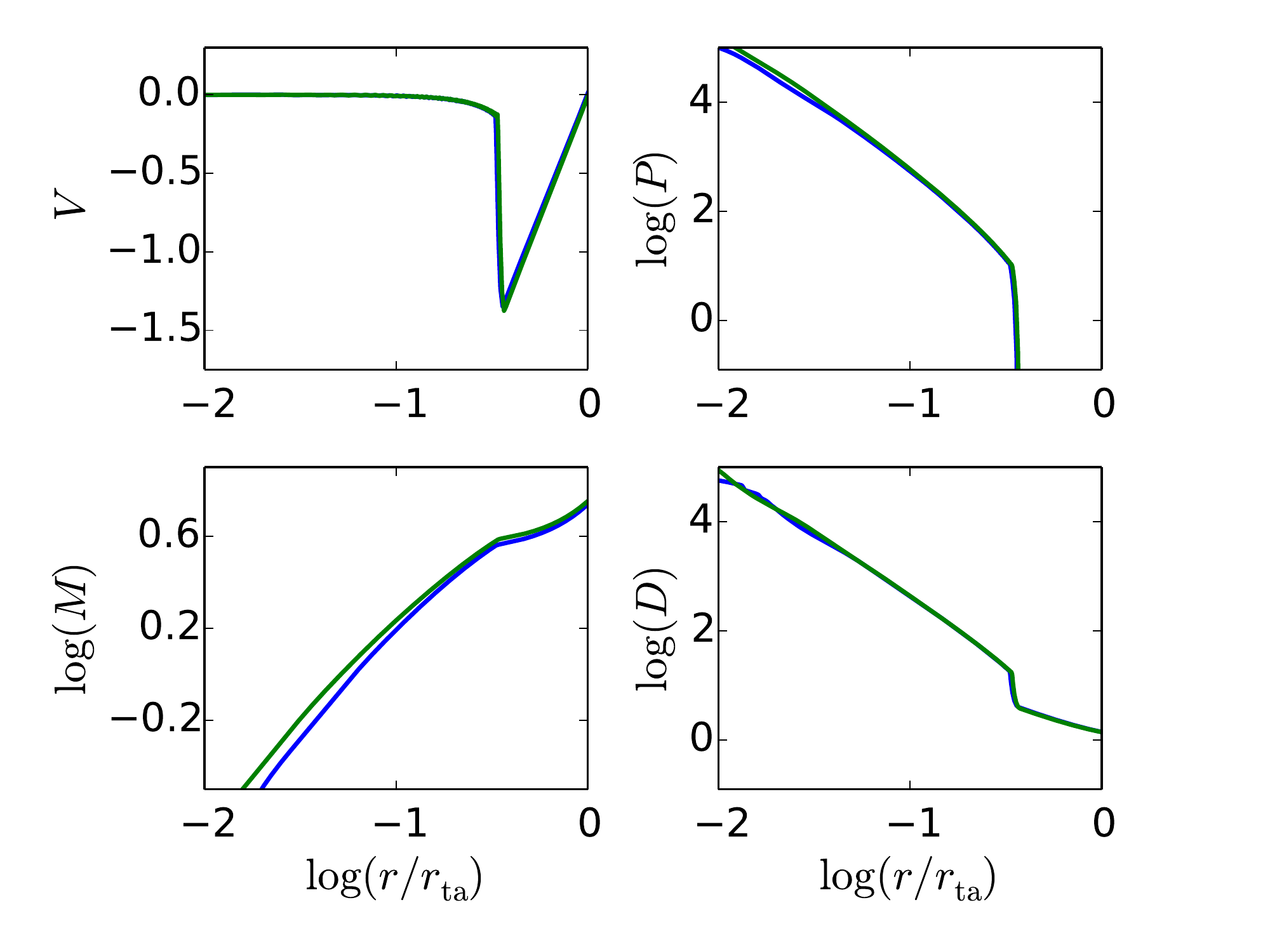}
\caption{Our simulation reproduces the results of \citet{thoul95} for self-similar fluid collapse with a Gaussian initial overdensity profile $\delta_{i}(r)=\delta_{i}(0)e^{-r^{2}/r_{i}^{2}}$, where $r_i$ is a scale radius. Top: all shells follow the same trajectory in non-dimensionalized coordinates. Bottom: non-dimensionalized velocity, pressure, mass, and density profiles at two snapshots in time (blue and green lines). These profiles agree with our analytic results.}
\label{fig:2.1}
\end{figure}

We perform several tests of our code. First, we initialise $N=1000$ shells in an EdS universe with a Gaussian initial overdensity profile $\delta_{i}(r) = [\rho_{i}(r) - \bar{\rho_{i}}]/\bar{\rho_{i}} = \delta_{i}(0)e^{-r^2/r_{i}^{2}}$, where $r_i$ is a scale radius, following \citet{thoul95}. Note that for $r \gg r_{i}$, $\delta M \rightarrow$\ const, so $\delta M/M \propto M^{-1}$ and this is equivalent to an $\epsilon =1$ mass perturbation. The resulting non-dimensionalized shell trajectories and fluid variable profiles (Figure \ref{fig:2.1}) agree with the results from Figures 3 and 4 of \citet{thoul95} and with the analytic self-similar solution for $\epsilon=1$ in \citet{bertschinger85}. We also reproduce Figures 6 and 7 of \citet{thoul95} for collapse from initial conditions generated by an $n=-2$ power-law power spectrum in an EdS universe. More generally, we have compared our hydrodynamic code to self-similar calculations (\S\ref{sect:self-similar}) for arbitrary $\epsilon$, finding excellent agreement (for example, see Figure \ref{fig:2.5}). The integrals of motion (equations \ref{eq:fluid_integrals1}, \ref{eq:fluid_integrals2}) are conserved to high precision.

\subsection{Realistic Initial Conditions} \label{sect:realistic}

\begin{figure}
\centering
\includegraphics[width=1.0\columnwidth]{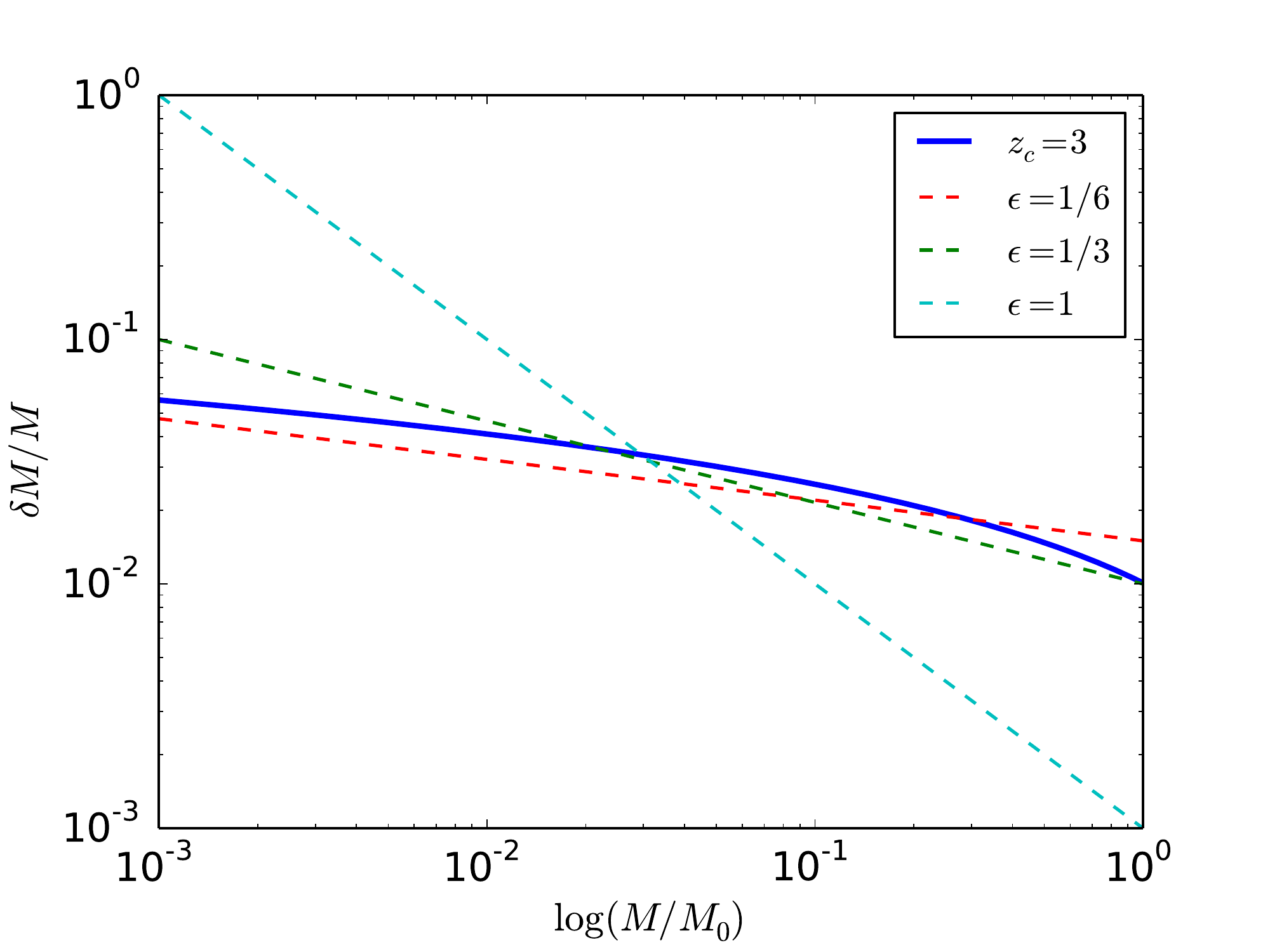} 
\caption{The initial overdensity profile (blue) generated from a \citet{wechsler02} halo mass accretion history using the linear perturbation technique from \citet{lu06}. The profile is initialised at redshift $z_{i} = 500$ with $z_{c}=3$. Dashed lines illustrate scale-free initial conditions, and $M_{0} = 10^{12}\ \msun$ is a reference mass.}
\label{fig:2.2}
\end{figure}

We now describe our method for simulating dark matter halo collapse with realistic initial conditions in a $\Lambda$CDM universe with \{$\Omega_{\Lambda},\Omega_{\text{m}}\}=\{0.721,0.279\}$. To generate realistic initial conditions, we follow \citet{lu06} by using linear perturbation theory to construct initial overdensity profiles that correspond to realistic halo mass accretion histories. In particular, we use the universal form for halo mass accretion histories found in \citet{wechsler02} as a fit to numerical simulations: 
\begin{equation}M(z) = M_{0}\ \text{exp}\Big[\frac{-S}{1+z_{c}}\big(\frac{1+z}{1+z_{f}}-1\big)\Big], \label{eqn:2.2}\end{equation}
where the parameter $z_{c}$ is the redshift at which the mass accretion rate $\text{d}(\text{log}M)/\text{d}(\text{log}a)$ falls below a critical value $S=2$ and effectively transitions from `fast' to `slow' accretion. A spherical shell collapses when its average linear overdensity reaches a critical value $\delta_{c} \approx 1.686$, so for a given $z_{c}$ we can construct the initial overdensity profile that gives rise to the mass accretion history in equation \ref{eqn:2.2} as it evolves in a $\Lambda\text{CDM}$ universe by choosing a radius partition $r_{1} < \hdots\ r_{j} < \hdots\ r_{N}$ and solving
\begin{equation}r_{j}(M) = \Big{\{} \frac{3M}{4\pi \bar{\rho}(z_{i})[1+\delta_{i}(M)]}\Big{\}}^{1/3} \label{eqn:2.3}\end{equation}
for $M(r)$. Here, $\delta_{i}(M) = 1.686\ D(z_{i})/D(z(M))$, where $D(z)$ is the linear growth factor (fit by \citealt{carroll92}) and $z(M)$ is given by inverting equation \ref{eqn:2.2}. Furthermore, the mean mass density at redshift $z_{i}$ is $\bar{\rho}(z_{i}) = \rho_{\text{crit,0}}\Omega_{\text{m}}(1+z_{i})^3$, where $\rho_{\text{crit,0}}$ is the critical density of the universe at $z=0$. We choose $N=4000$ equally spaced shells for our simulation; an example of the initial overdensity profile generated for a $M_{0} = 10^{12}\ \msun$ halo with $z_{c} = 3$ and $z_{i} = 500$ is shown in Figure \ref{fig:2.2}. Our conclusions are insensitive to the values of $M_{0}$, $z_{c}$, and $z_{i}$, as long as $z_{i} >> z_{c}$ so that the linear perturbation technique is valid.

\begin{figure}
\centering
\includegraphics[width=1.0\columnwidth]{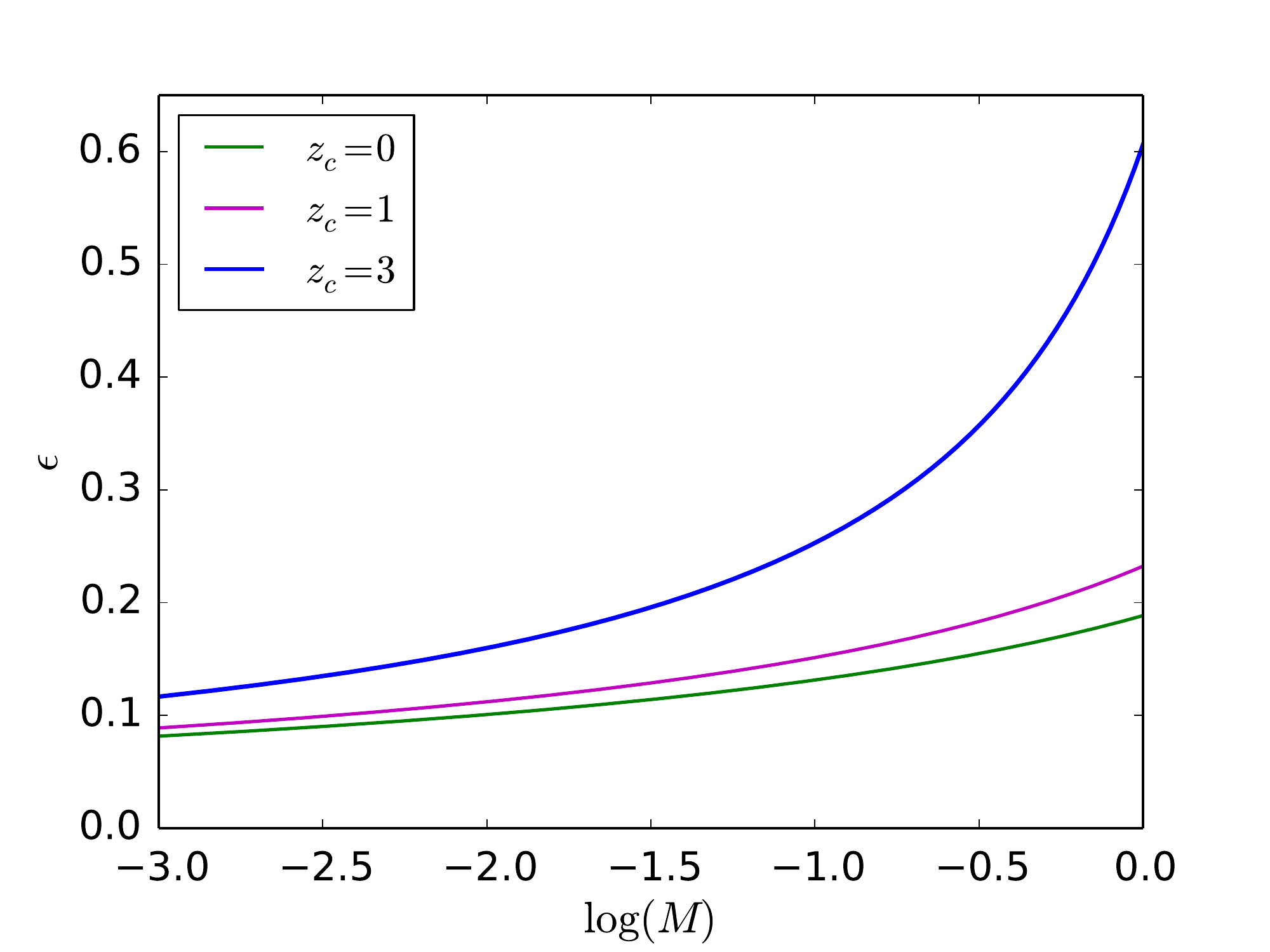} \caption{$\epsilon(M)\equiv -{\text{d}(\text{log}[\delta M/M])}/{\text{d}(\text{log}M)}$ for three halos that collapse from realistic initial conditions at $z_{i}=500$. The realistic initial conditions are not scale free ($\epsilon \neq \text{const}$), so there is no reason to expect a power-law pseudo phase space density profile. $M$ is the enclosed mass at $z=0$ normalised to the mass enclosed at the shock radius. We choose $z_{c}=3$ (blue) as our fiducial case throughout this paper.}
\label{fig:2.6}
\end{figure}

In Figure \ref{fig:2.6}, we show the $\epsilon(M)$ profiles for $z_{c} = 0$, $1$, and $3$, where $\epsilon(M) \equiv -{\text{d}(\text{log}[\delta M/M])}/{\text{d}(\text{log}M)}$ (motivated by $\delta M/M \propto M^{-\epsilon}$ in the scale-free case) and $M$ is evaluated at $z=0$. Though $\epsilon \approx 1/6$ in the inner regions of the collapsed halos (as we expect from the $n_{\text{eff}}\sim -2.5$ power-law index of the power spectrum on galactic scales), $\epsilon(M)$ varies substantially over each halo profile. In this paper, we choose $z_{\rm c}=3$ as our fiducial case; this value is characteristic of halos with $M_{\rm halo} \sim 10^{12} \, M_{\odot}$ at $z=0$ (e.g., see Figure 11 of \citet{wechsler02}). Note that $z_{\rm c}=3$ is similar to the value chosen in \citet{shapiro06-entropy}, where a comparable spherical collapse model is applied. We obtain very similar results for halos with different values of $z_{\rm c}$.

\begin{figure}
\centering
\includegraphics[width=1.0\columnwidth]{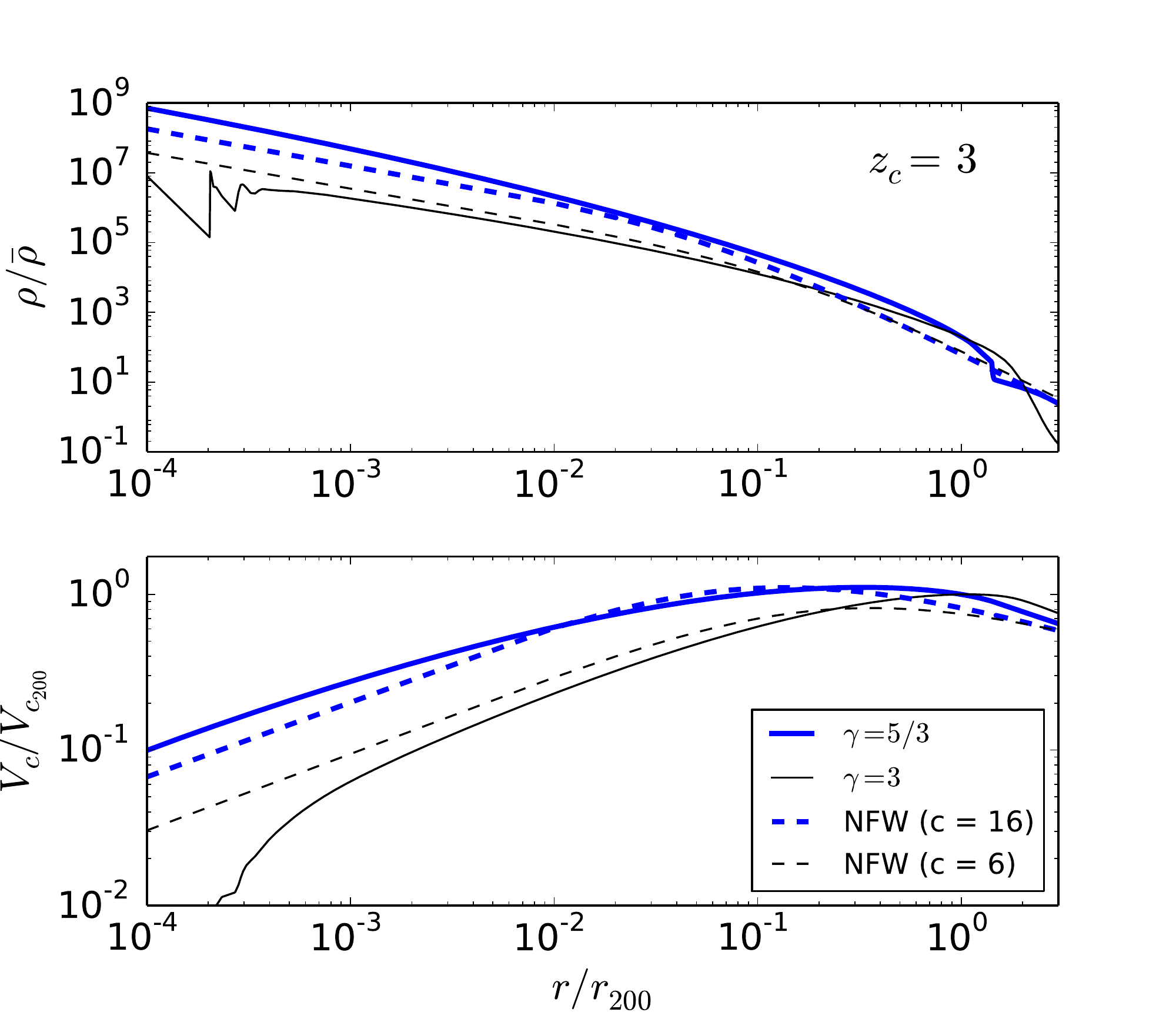} \caption{Density and circular velocity profiles at $z=0$ from our hydrodynamic simulation with realistic initial conditions (solid curves), along with the best-fitting NFW profiles (dashed lines). $r_{200}$ is the virial radius and $V_{c_{200}}$ is the circular velocity at the virial radius. Varying the adiabatic index $\gamma$ models the effects of velocity anisotropy.}
\label{fig:2.3}
\end{figure}

To simulate halo collapse in a  $\Lambda$CDM universe, we model the effective gravitational acceleration of the $j$th mass shell as
\begin{equation}\frac{\text{d}^{2}r_{j}}{\text{d}t^{2}} = H_{0}^{2}\Omega_{\Lambda}r_{j} - \frac{GM_{j}}{r_{j}^{2}}, \label{eqn:2.4}\end{equation}
where $r_{j}$ is the radius of the shell and $M_{j}$ is the mass it encloses. The initial velocity profile is purely radial and corresponds to the unperturbed Hubble flow at redshift $z_{i}$. Figure \ref{fig:2.3} shows the $z=0$ density and circular velocity profiles for collapse from the realistic initial conditions of Figure \ref{fig:2.2}. Our results are fit reasonably well by the corresponding NFW profiles, which shows that the universality in halo density and circular velocity profiles can be understood as the result of realistic initial conditions\footnote{The rough agreement between our simulation results and the NFW profiles is similar to the result in \cite{shapiro04-halo}, but we use a different method to generate initial conditions.}. 

The NFW concentration parameter that fits our results ($c=16$) is large for a $M_{0} = 10^{12}\ \msun$ halo (see \citealt{ludlow16}). We therefore explore the effects of varying the adiabatic index $\gamma$ in order to crudely model velocity anisotropy, which is a crucial effect in shaping density profiles with lower values of $c$ \citep{lu06}. We find that fluid collapse with $\gamma=3$ -- i.e., a radial velocity dispersion -- yields a much more reasonable concentration parameter $c \sim 6$. 
Varying $\gamma$ shell-by-shell can in principle account for velocity anisotropy; the large value we obtain for $c$ is therefore an artefact of our fluid model. Since our focus is on the entropy profile, we sidestep this issue in this paper.

The density profile in Figure \ref{fig:2.3} differs from an NFW profile in detail, since our 1D fluid calculation only approximates 3D collisionless dynamics; however, the inner and outer slopes of the density profile are consistent with the NFW values. Moreover, we will see that the entropy profiles produced by our 1D model correspond to pseudo phase space density profiles that closely resemble those found in N-body simulations. This suggests that the detailed behaviour of the density profile is largely irrelevant for understanding whether the pseudo phase space density is a power law.

\begin{figure}
\centering
\includegraphics[width=1.0\columnwidth]{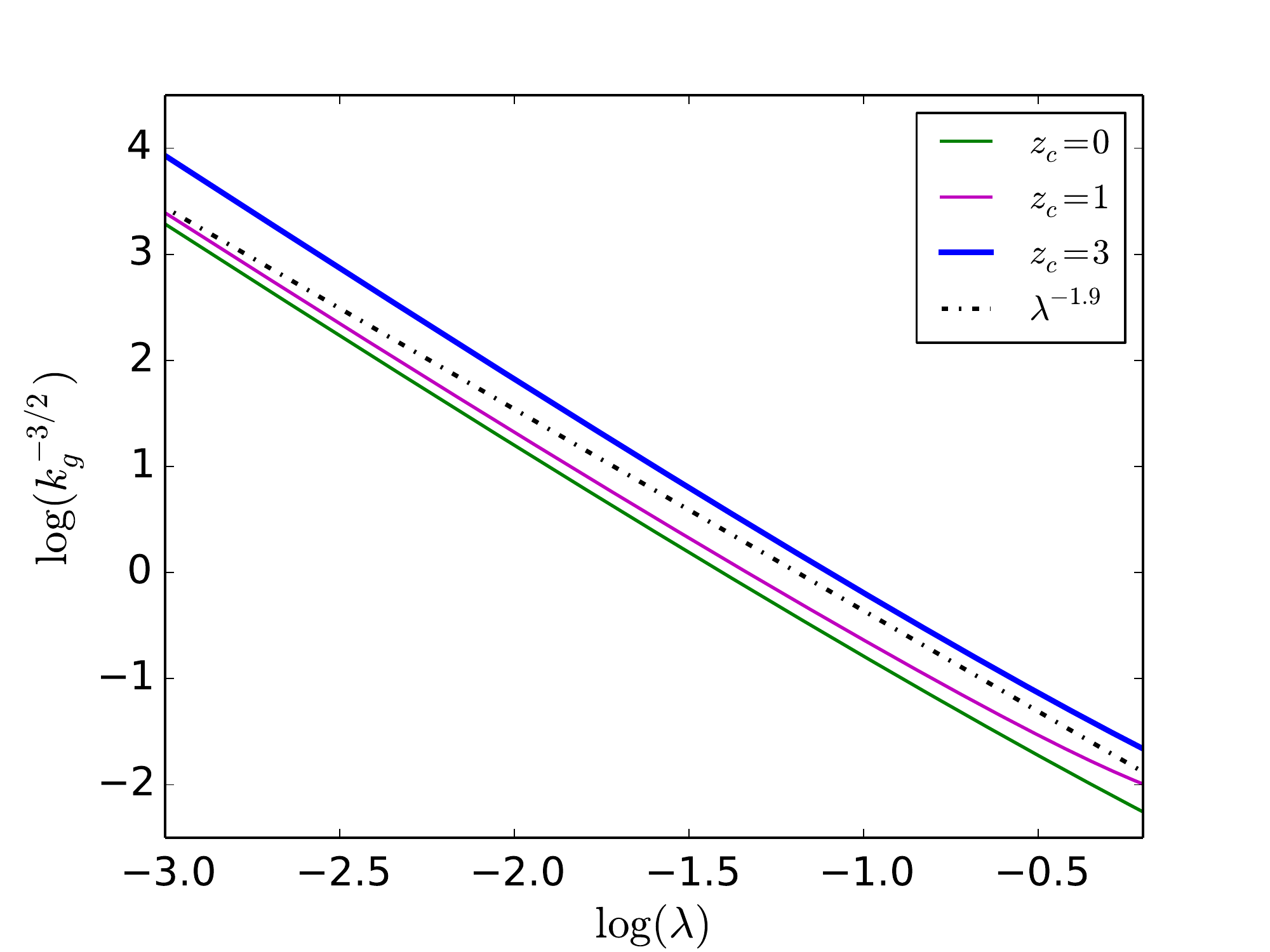} \caption{Pseudo phase space density profiles $Q(\lambda) \propto k_{g}^{-3/2}$ at $z=0$ for three halos that collapse from realistic initial conditions at $z_{i} = 500$ ($\lambda$ is the radius normalised to the fluid shock). The profiles seem to follow a universal power law that agrees reasonably well with the fit from N-body simulations. However, these profiles are not true power laws; see Figure \ref{fig:2.7}.}
\label{fig:2.4}
\end{figure}

As discussed above, we can construct the fluid analogue of the pseudo phase space density using $Q(r) \propto k_{g}^{-3/2}$. Figure \ref{fig:2.4} shows the $z=0$ pseudo phase space density profiles that correspond to the three initial conditions from Figure \ref{fig:2.6}. The profiles agree reasonably well with the power law favoured by N-body simulations over the dynamic range shown; \citet{shapiro06-entropy} reach a similar result and conclude that the universality in $Q(r)$ profiles can be understood as the result of realistic initial conditions, even in the context of simple spherical collapse models. In particular, they argue that a power-law fit to $Q(r)$ very similar to those obtained from N-body simulations results from a realistic time-varying mass accretion history in a $\Lambda$CDM universe. However, we will find that the $Q(r)$ profiles produced by our spherical collapse model deviate significantly from power laws over the dynamic range relevant to N-body simulations. In particular, the logarithmic derivatives of the curves in Figure \ref{fig:2.4} are not constant functions (see Figure \ref{fig:2.7}). Thus, by carefully examining the logarithmic derivatives of our pseudo phase space density profiles, we arrive at different conclusions than \citet{shapiro06-entropy}.

\section{Results} 
\label{sect:results}

\subsection{Self-similar Fluid Collapse}
\label{sect:power_law_results}  

\begin{figure}
\includegraphics[width=1.0\columnwidth]{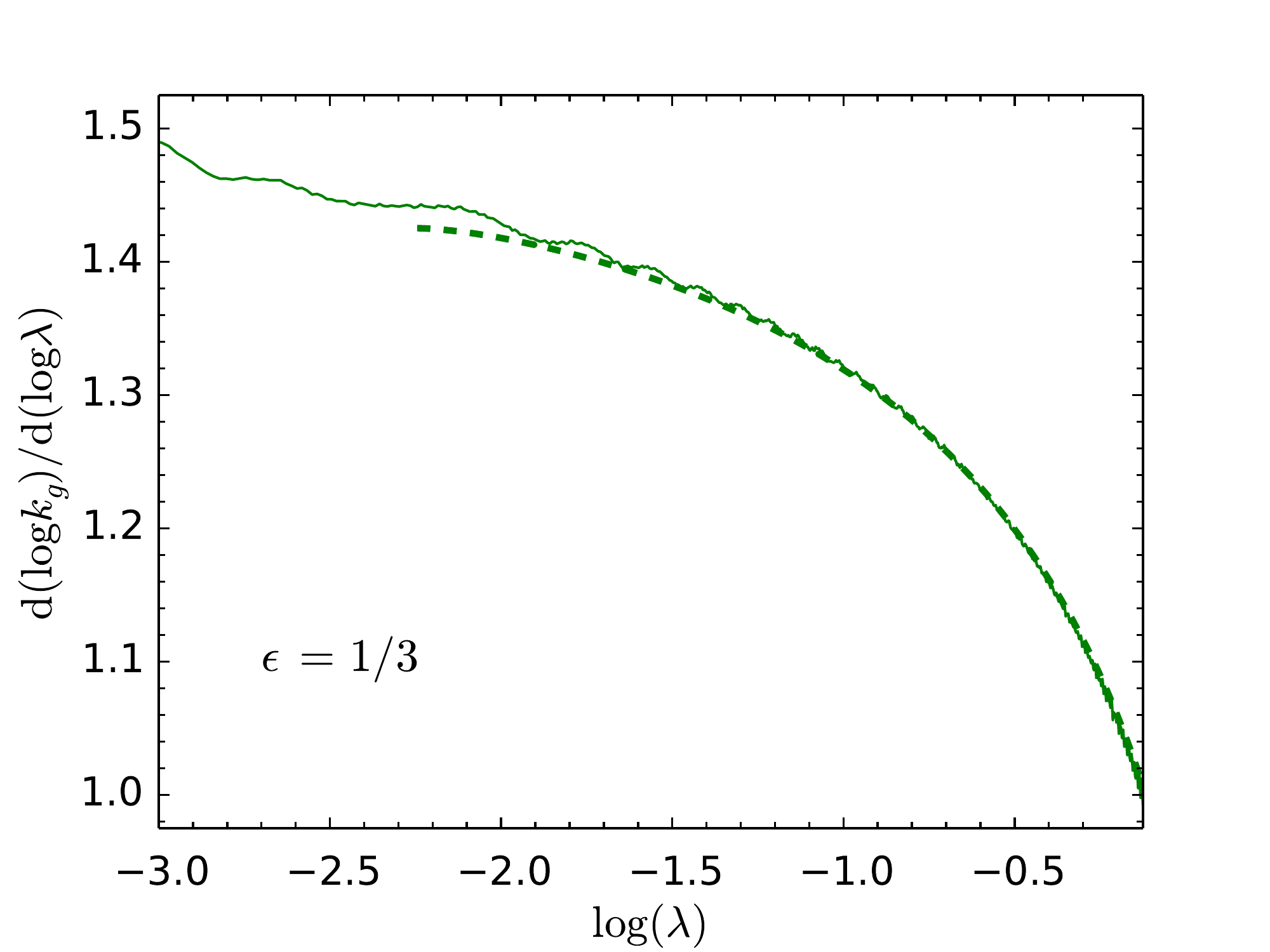}
\includegraphics[width=1.0\columnwidth]{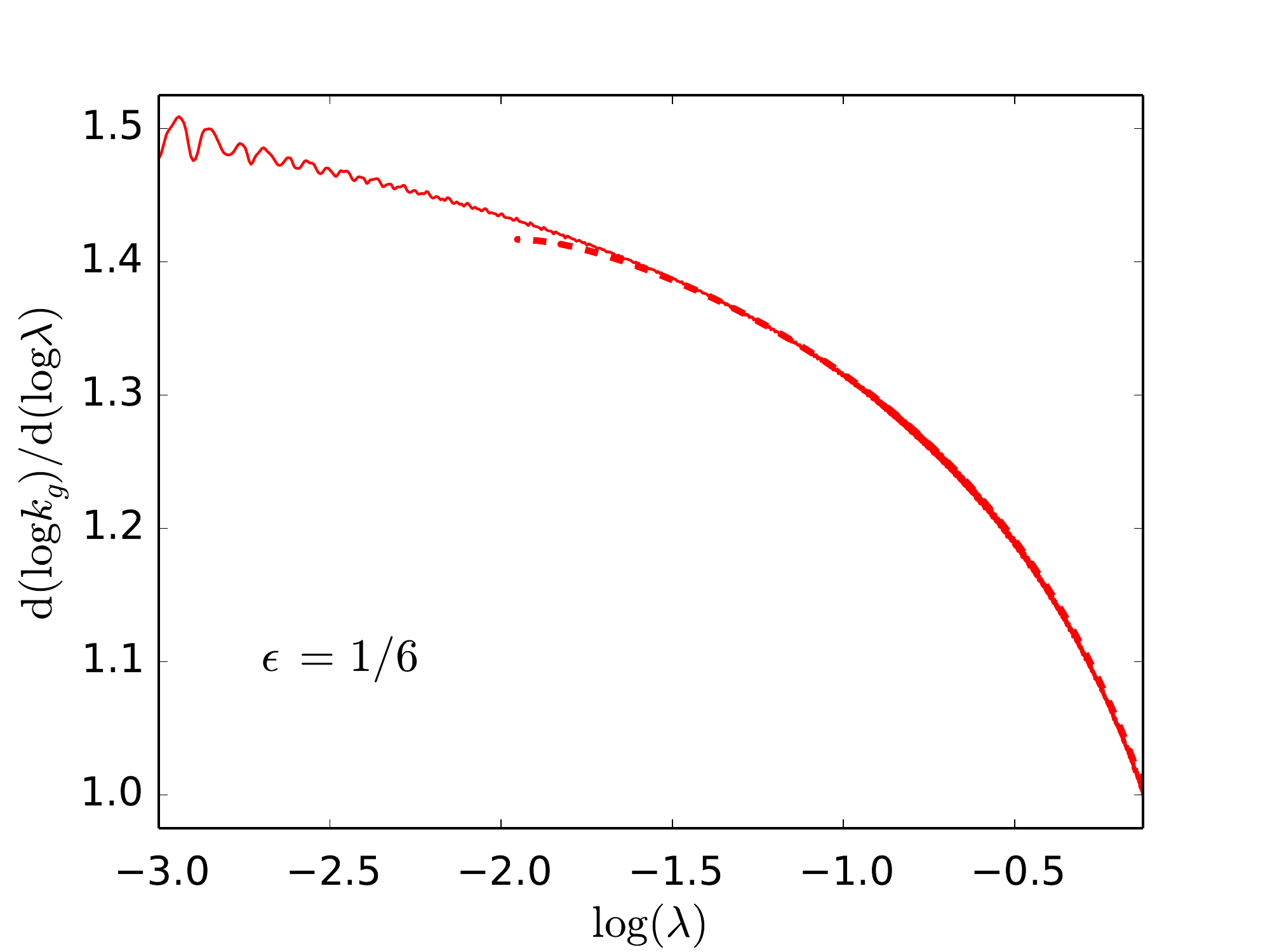}
\caption{Logarithmic derivatives of the $z=0$ entropy profiles from our analytic calculation (dashed lines) and our simulation (solid lines) for collapse from scale-free initial conditions $\delta M/M \propto M^{-\epsilon}$, with $\epsilon=1/3$ (top) and $\epsilon=1/6$ (bottom). $\lambda$ is the radius normalised to the fluid shock. Note that the physical shock radius becomes too small to resolve using our analytic method for $\log(\lambda) \lesssim -2.0$.}
\label{fig:2.5}
\end{figure}

We first examine the entropy profiles for collapse from scale-free initial conditions in an EdS cosmology, which results in self-similar collapse (\S\ref{sect:self-similar}). These calculations provide much of the physical insight required to understand more realistic entropy profiles, which we consider in \S\ref{sect:results_realistic}. We study scale-free collapse with $\epsilon=1/3$ (green lines) and $1/6$ (red lines) in detail in Figures \ref{fig:2.5}--\ref{fig:2.10}. We choose these values for detailed study because they correspond to the central ($M\rightarrow 0$) and average values of $\epsilon(M)$ for our fiducial realistic profile with $z_{c}=3$ (Figure \ref{fig:2.6}; note that $\epsilon=1/6$ is more representative for lower values of $z_{c}$). Just as for more realistic initial conditions, scale-free collapse yields power-law-like pseudo phase space density (PPSD) profiles with promising straight-line fits on log-log plots. We now list our main results.

{\it Entropy is a power-law function of mass, but not of radius.} From equation \ref{eq:fluid_integrals2}, we expect the gas entropy to be an exact power-law function of mass, since entropy is a Lagrangian conserved quantity in an adiabatic fluid. In particular, for scale-free collapse in an EdS universe,
\begin{equation}\frac{\text{d}(\text{log}k_{g})}{\text{d}(\text{log}M)} = -\frac{6}{n}\frac{\alpha+\gamma-2}{2-3\alpha}, \label{eqn:2.8}\end{equation}
where $n=1,2,3$ for planar, cylindrical, and spherical perturbations (see, e.g., \citealt{chuzhoy00}); for $n=3$, this reproduces equation \ref{eq:fluid_integrals2}. The lower panel of Figure \ref{fig:2.8} shows that our fluid simulation recovers this result. The fact that this integral of motion is conserved to high precision demonstrates the accuracy of our hydrodynamic simulation.

However, entropy is \emph{not} a power-law function of radius. In Figure \ref{fig:2.5}, we show the logarithmic entropy derivative $\text{d}(\text{log}k_{g})/\text{d}(\text{log}\lambda)$ from our analytic calculation (\S\ref{sect:self-similar}) and our hydrodynamic simulation (\S\ref{sect:hydro}) in an EdS cosmology. The close agreement between the profiles lends confidence to the the fidelity of our simulation. Thus, we only present simulation results in Figures \ref{fig:2.8} and \ref{fig:2.7}. The key takeaway from Figure \ref{fig:2.5} is that $\text{d}(\text{log}k_{g})/\text{d}(\text{log}\lambda)$ is not constant, so the gas entropy is {\it not} a power-law function of radius, even for self-similar collapse! 

{\it Departures from hydrostatic equilibrium (HSE) are responsible for the deviation from a radial power law.} Using $k_{g} = P D^{-\gamma}$ and equation \ref{eq:fluid_self_similar3}, we obtain
\begin{equation}
\frac{\text{d}(\text{log}k_{g})}{\text{d}(\text{log}\lambda)} = \frac{\lambda[2(2-\alpha)-2\gamma]}{V-\alpha \lambda},  \label{eqn:2.7}
\end{equation}
where $\lambda$ is the radius scaled to the fluid shock and $\alpha$ is the power-law index of characteristic scale radii as functions of time (equation \ref{eqn:1.3}). Equation \ref{eqn:2.7} implies that, if the halo is in hydrostatic equilibrium ($V=0$), the entropy profile is an \emph{exact} power law. Otherwise, for a nonzero radial velocity $V(\lambda)$, the entropy only attains a constant logarithmic slope asymptotically\footnote{A power law is also obtained if $V \propto \lambda$. In practice, this only occurs in a neighbourhood near $\lambda=0$, where the boundary condition $V(0)=0$ guarantees that the linear term in a Taylor expansion is a good approximation.}, as $\lambda \rightarrow 0$, recalling the boundary condition $V(0)=0$. Deviations from hydrostatic equilibrium therefore cause the power-law slope of the entropy profile to run systematically with radius. 

It is easy to see that violations to hydrostatic equilibrium should be sufficient to cause the radial power-law index of the entropy profile, $\alpha_{\lambda} \equiv \text{d}(\text{log}k_{g})/\text{d}(\text{log}\lambda)$, to `roll'. For $\alpha_{\lambda}$ to be constant, we require $V \ll \alpha \lambda$. However, at the shock, $V \sim \mathcal{O}(\alpha \lambda)$, where $\alpha \lambda$ is the non-dimensionalized shock speed. $V$ slowly declines throughout the halo as fluid elements are decelerated by the excess pressure gradient, until $V(0)=0$. We note that {\it none} of the fluid variables for self-similar collapse are power-law functions of radius, even though they might appear to be straight lines on a log-log plot (e.g., see Figure \ref{fig:2.1}). These quantities only \emph{asymptotically} approach power laws as $\lambda \rightarrow 0$. In this limit, $V\rightarrow V_{0}\lambda$, $D \propto \lambda^{\delta}$, $P \propto \lambda^{\eta}$, and $M \propto \lambda^{\delta + \eta}$, where $V_{0}=\{4(6\epsilon-1)/(45\epsilon), 0\}$, $\delta=\{-3(3\epsilon +2)/(3\epsilon+7), - 9\epsilon/(3\epsilon+1)\}$, and $\eta = \{0, 2(1-6\epsilon)/(3\epsilon+1)\}$ for $\{\epsilon \le 1/6, \epsilon > 1/6\}$ respectively\footnote{Note the dichotomy between $\epsilon \le 1/6$, where the central density continuously increases with time, and $\epsilon > 1/6$, where the central density is constant (which can be seen from $V_{0}=0$).} \citep{chuzhoy00}.

{\it Convergence to HSE -- and to the asymptotic power-law entropy slope -- is slow.} We can understand this by examining the velocity profiles. In Figure \ref{fig:2.10}, we plot the quantity $\left|\text{log}[1-V/(\alpha \lambda)]\right|$ versus $\lambda$ (in the realistic case, $\alpha$ is determined as a function of $\lambda$ by using the realistic $\epsilon(\lambda)$ profile in equation \ref{eqn:1.3}). For the entropy profile to approach a power law, this quantity must approach 0 (for $V\rightarrow 0$) or a constant (for $V \rightarrow V_{0} \lambda$). Figure \ref{fig:2.10} shows that the velocity profiles approach their asymptotic values very slowly. In addition, the convergence to HSE is faster for larger values of $\epsilon$, corresponding to more sharply peaked initial profiles. The slow approach to hydrostatic equilibrium, and the trend with $\epsilon$, are consistent with the lower panel of Figure \ref{fig:2.7}. In particular, the entropy profiles in Figure \ref{fig:2.7} approach their asymptotic central values (shown by the dashed lines) very slowly (i.e., at small $\lambda$), and the convergence is faster (i.e., it occurs at larger $\lambda$) for larger values of $\epsilon$.

Note that the close similarity between the $\text{d}(\text{log}k_{g})/\text{d}(\text{log}\lambda)$ profiles for $\epsilon=1/3$ and $1/6$ is coincidental. The asymptotic slope for $\epsilon=1/6$ is steeper than that for $\epsilon=1/3$ (as shown by the red and green dashed lines in Figure \ref{fig:2.7}), but the $\epsilon=1/6$ profile approaches its asymptotic value more slowly due to the slower convergence to HSE. In the radial range we study, these two effects happen to cancel. The accidental nature of this cancellation is highlighted by the differing $\text{d}(\text{log}k_{g})/\text{d}(\text{log}\lambda)$ profiles for $\epsilon=2/3$ (purple) and $\epsilon=1$ (light blue) in the bottom panel of Figure \ref{fig:2.7}.

{\it The convergence rate is linked to the accretion rate.} Deviations from hydrostatic equilibrium are related to the mass accretion rate. Given sufficient time, an isolated self-gravitating object will settle into perfect hydrostatic equilibrium. Mass accretion acts as a forcing function that promotes mass inflow ($V \ne 0$) and deepening of the potential well; the slower the mass accretion rate, the weaker the nature of this gravitational forcing. Since $M \propto t^{2/(3\epsilon)}$, larger values of $\epsilon$ correspond to smaller mass accretion rates, and thus smaller departures from HSE. This leads to faster convergence to asymptotic power-law entropy slopes for larger values of $\epsilon$, as we noted above. 

These results lead to the strong suspicion that, barring some fortuitous cancellation, entropy profiles for more realistic scenarios should not be power-law functions of radius. We now examine the more realistic case.

\subsection{Realistic Initial Conditions}
\label{sect:results_realistic} 

We consider the realistic initial conditions and cosmology described in \S\ref{sect:realistic}. There are now two features that break self-similarity. First, the transition from an EdS to a $\Lambda$CDM cosmology removes the scale-free nature of Hubble expansion (for instance, the expansion rate and the mean density are no longer power-law functions of time). In addition, as we have seen in \S\ref{sect:realistic}, the mass excess $\delta M/M$ is no longer a power law; instead, it is a rolling power law with an index $\epsilon(M)$ (see Figure \ref{fig:2.6}). As noted above, we choose $z_{c}=3$ as our fiducial case, although we obtain very similar results for initial conditions with different values of $z_{c}$. Our results are as follows.

\begin{figure}
\includegraphics[width=1\columnwidth]{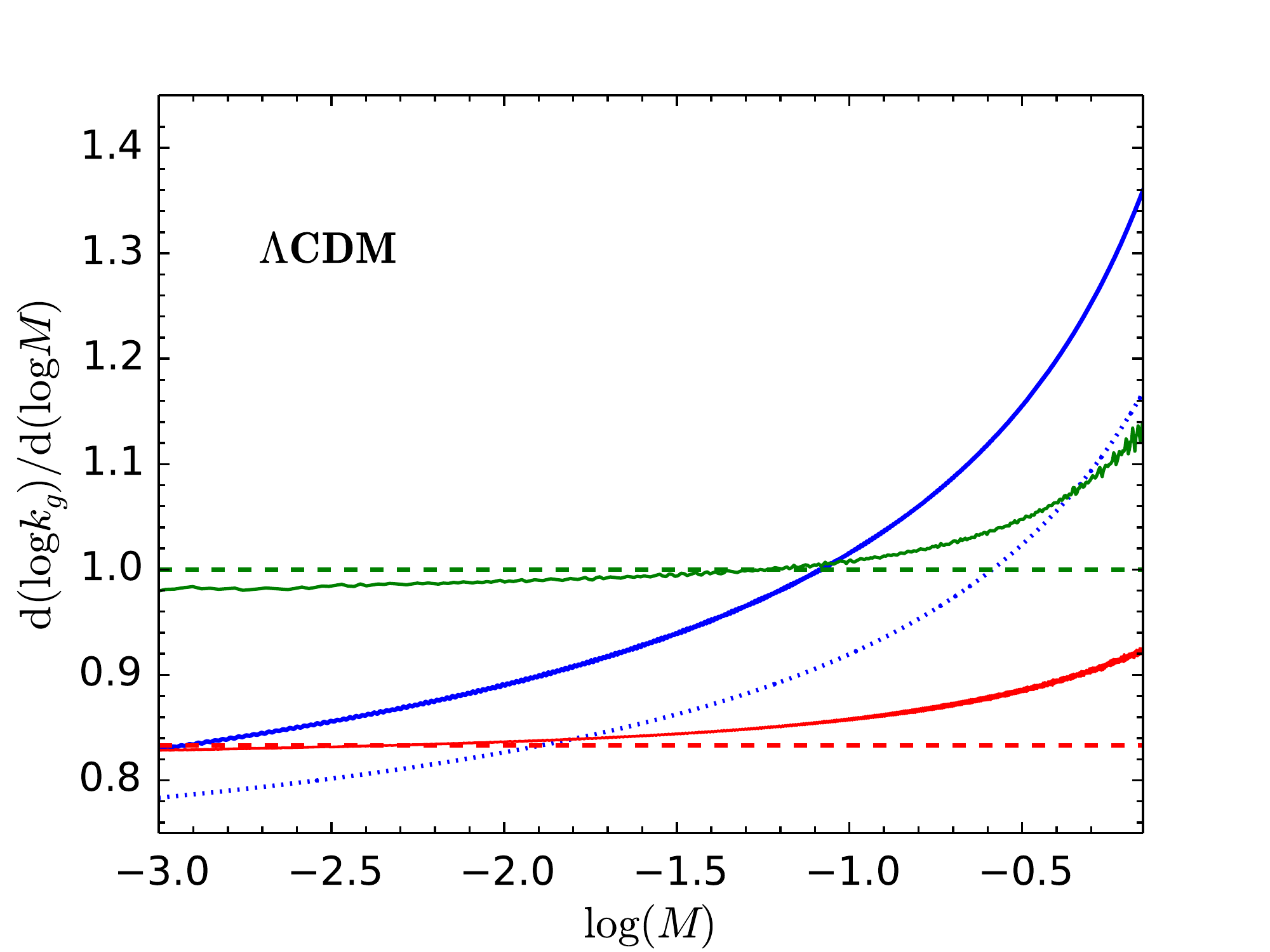}
\includegraphics[width=1\columnwidth]{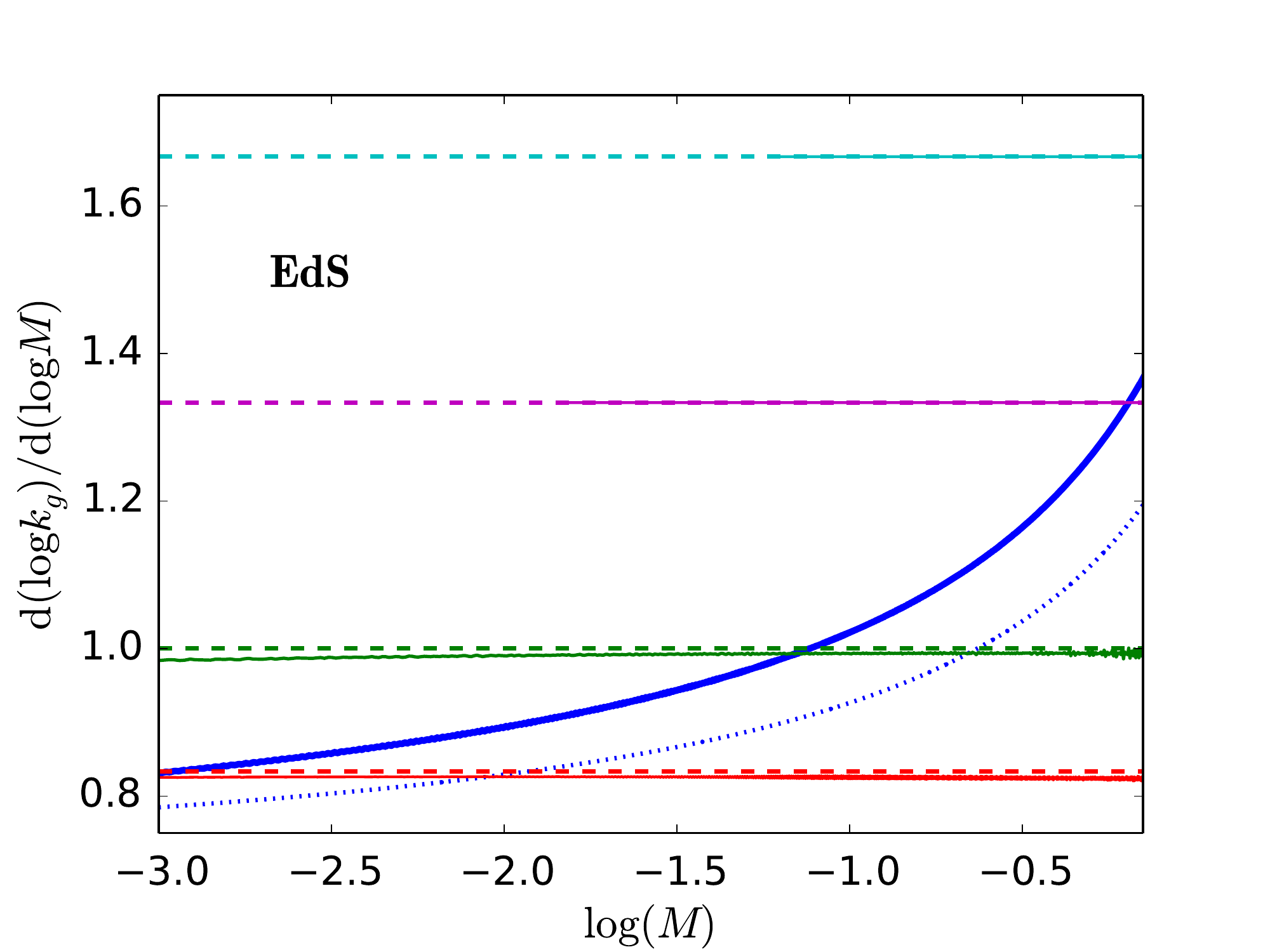}
\caption{$\text{d}(\text{log}k_{g})/\text{d}(\text{log}M)$ at $z=0$ for collapse from realistic initial conditions (blue) and scale-free initial conditions with $\epsilon=1/3$ (green) and $\epsilon=1/6$ (red), in a $\Lambda$CDM (top) and EdS (bottom) cosmology. Analytic results for $\epsilon=2/3$ (purple) and $\epsilon=1$ (light blue) are shown in the bottom panel, and the blue dotted line is the approximation from equation \ref{eqn:2.9} for the realistic case. The dashed lines mark the analytic power-law slope for each scale-free profile (equation \ref{eqn:2.8}).}
\label{fig:2.8}
\end{figure}

{\it Entropy is no longer a power-law function of mass.} Since entropy is a power-law function of mass for self-similar collapse, we first check whether this remains true for more realistic initial conditions. Figure \ref{fig:2.8} shows that $k_{g}(M)$ is \emph{not} a power law. Instead, the power-law index of the entropy profile as a function of mass, $\alpha_{M} \equiv {\text{d}(\text{log}k_{g})}/{\text{d}(\text{log}M)}$, rolls with mass in both an EdS and a $\Lambda$CDM cosmology. Since the results are similar in either cosmology, this behaviour must arise from the scale-dependent initial conditions. Indeed, we can largely reproduce the realistic $\alpha_{M}$ profile by substituting the instantaneous value of $\alpha(\epsilon)$ (equation \ref{eqn:1.3}) into equation \ref{eqn:2.8}, which gives
\begin{equation}\frac{\text{d}(\text{log}k_{g})}{\text{d}(\text{log}M)} \approx \epsilon(M) + \frac{2}{3} \label{eqn:2.9}\end{equation}
\noindent for spherical collapse with $\gamma=5/3$. This relation can be understood intuitively from the argument in \S\ref{sect:fluid}: the gas entropy $k_{g} \propto T/\rho^{2/3} \propto v_{\rm in}^{2}/\rho^{2/3} \propto ({GM/R})/\rho^{2/3} \propto (M t)^{2/3} \propto M^{2/3(1+3\epsilon/2)} \propto M^{\epsilon +2/3}$, where we have used $M\propto t^{2/(3\epsilon)}$ in the penultimate step. Equation \ref{eqn:2.9} is an approximate equality because $\alpha$ is only well defined for scale-free collapse. However, this relation seems quite robust; the $\alpha_{M}$ profile predicted using the realistic $\epsilon(M)$ profile in equation \ref{eqn:2.9} matches our simulation to better than 15$\%$ over the entire collapsed halo (see Figure \ref{fig:2.8}). In fact, this approximation largely reproduces the shape of the realistic $\alpha_{M}$ profile and differs mainly in its normalisation.

From equation \ref{eqn:2.9}, we see that the slowly rolling ${\text{d}(\text{log}k_{g})}/{\text{d}(\text{log}M)}$ profile is directly related to $\epsilon(M)$. The latter rolls slowly with mass (Figure \ref{fig:2.6}) because the excess mass $\delta M$ is an integral quantity that cannot change abruptly\footnote{Equivalently, the integral excess energy $\delta E$ is a slowly rolling power law. The excess energy and post-shock entropy of a shell are closely related conserved Lagrangian quantities.} if $\delta M \ll M$, as required for linear theory to be valid\footnote{Even a delta function overdensity at the origin leads to $\delta M/M \propto M^{-1}$ in the linear regime; this corresponds to the `secondary infall' model studied by \citet{bertschinger85}.}. Thus, $k_{g} \propto M^{\zeta}$, where the mean index $\zeta \approx 1$, with a maximum $\sim 20\%$ deviation on either end (see Figure \ref{fig:2.9}).

\begin{figure}
\includegraphics[width=1.0\columnwidth]{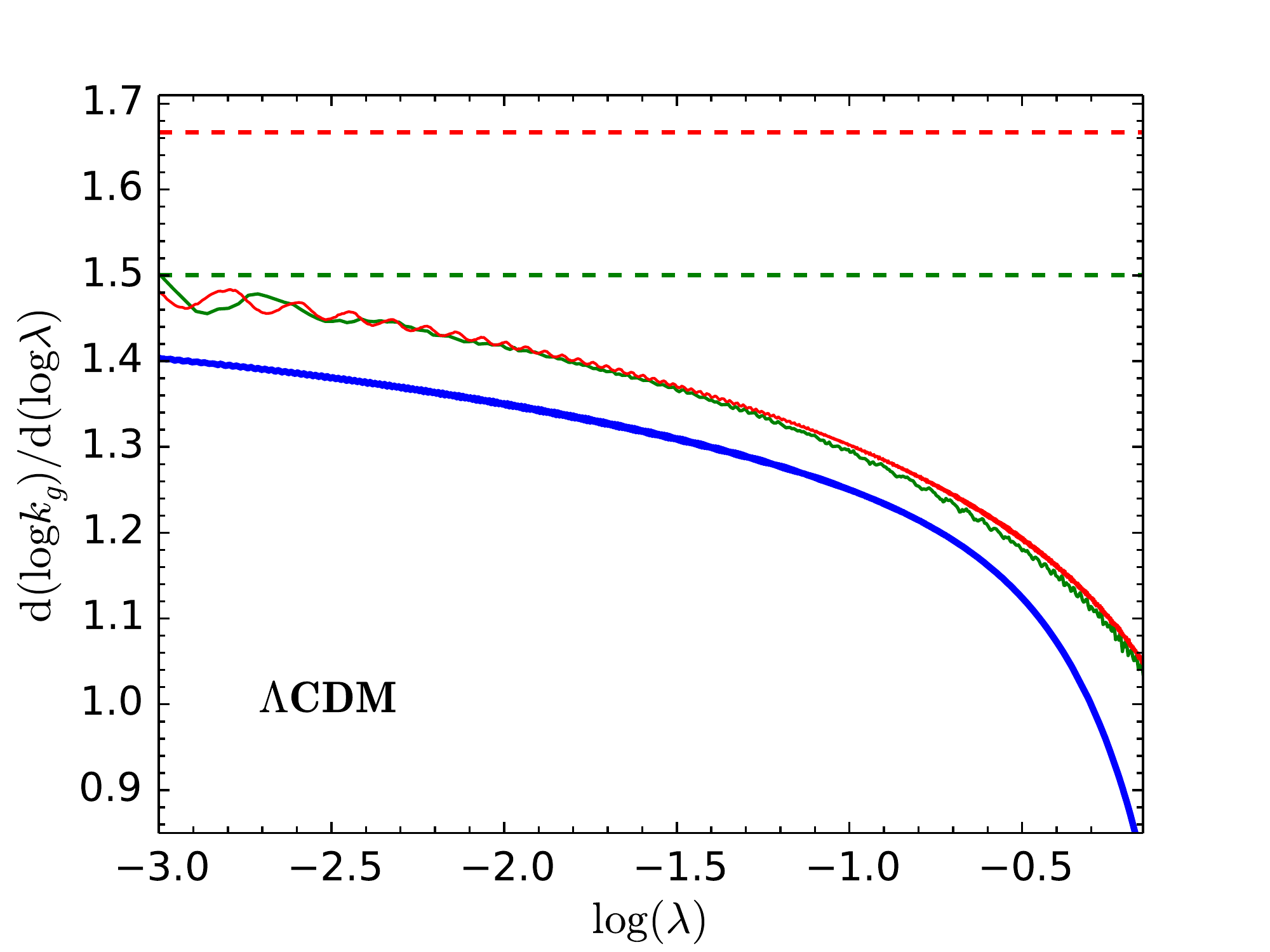}
\includegraphics[width=1\columnwidth]{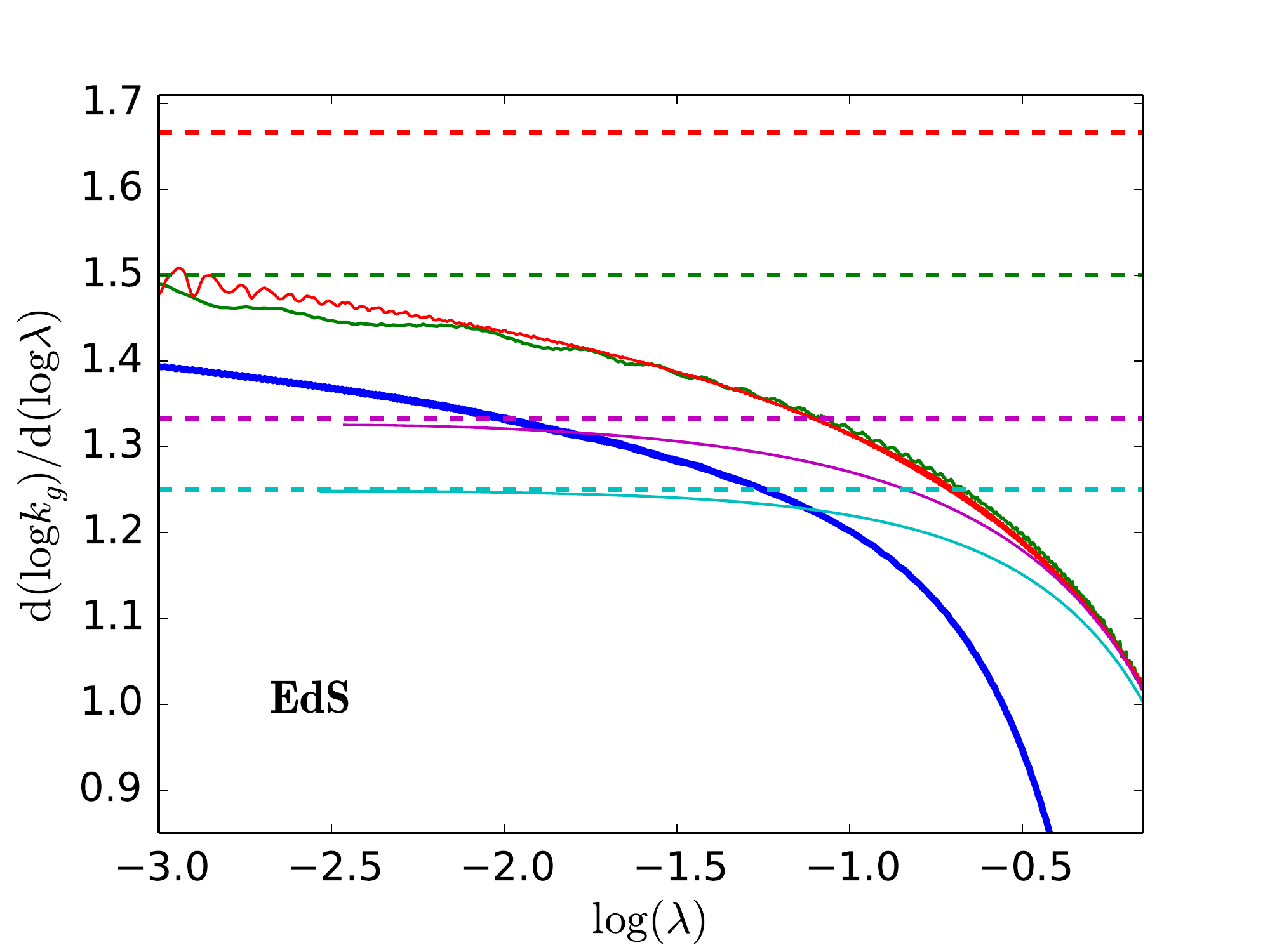}
\caption{Logarithmic derivatives of the $z=0$ entropy profiles for collapse from realistic initial conditions (blue) and scale-free initial conditions with $\epsilon=1/3$ (green) and $\epsilon=1/6$ (red), in a $\Lambda$CDM (top) and EdS (bottom) cosmology. Analytic results for $\epsilon=2/3$ (purple) and $\epsilon=1$ (light blue) are shown in the bottom panel. The dashed lines mark the asymptotic power-law slope for each scale-free profile (equation \ref{eqn:2.7}).}
\label{fig:2.7}
\end{figure}

{\it Entropy is not a power-law function of radius.} The previous results lead us to believe that the realistic entropy profile should not be a radial power law; Figure \ref{fig:2.7} shows that the radial power-law index of the entropy profile rolls as a function of $\lambda$, as we expect. The departure from a power law is somewhat larger than for self-similar collapse, since two effects now contribute: deviations from hydrostatic equilibrium, and the fact that the collapse is no longer scale free. This result strongly suggests that the PPSD is {\it not} a power law for realistic CDM halos.

\begin{figure}
\centering
\includegraphics[width=1\columnwidth]{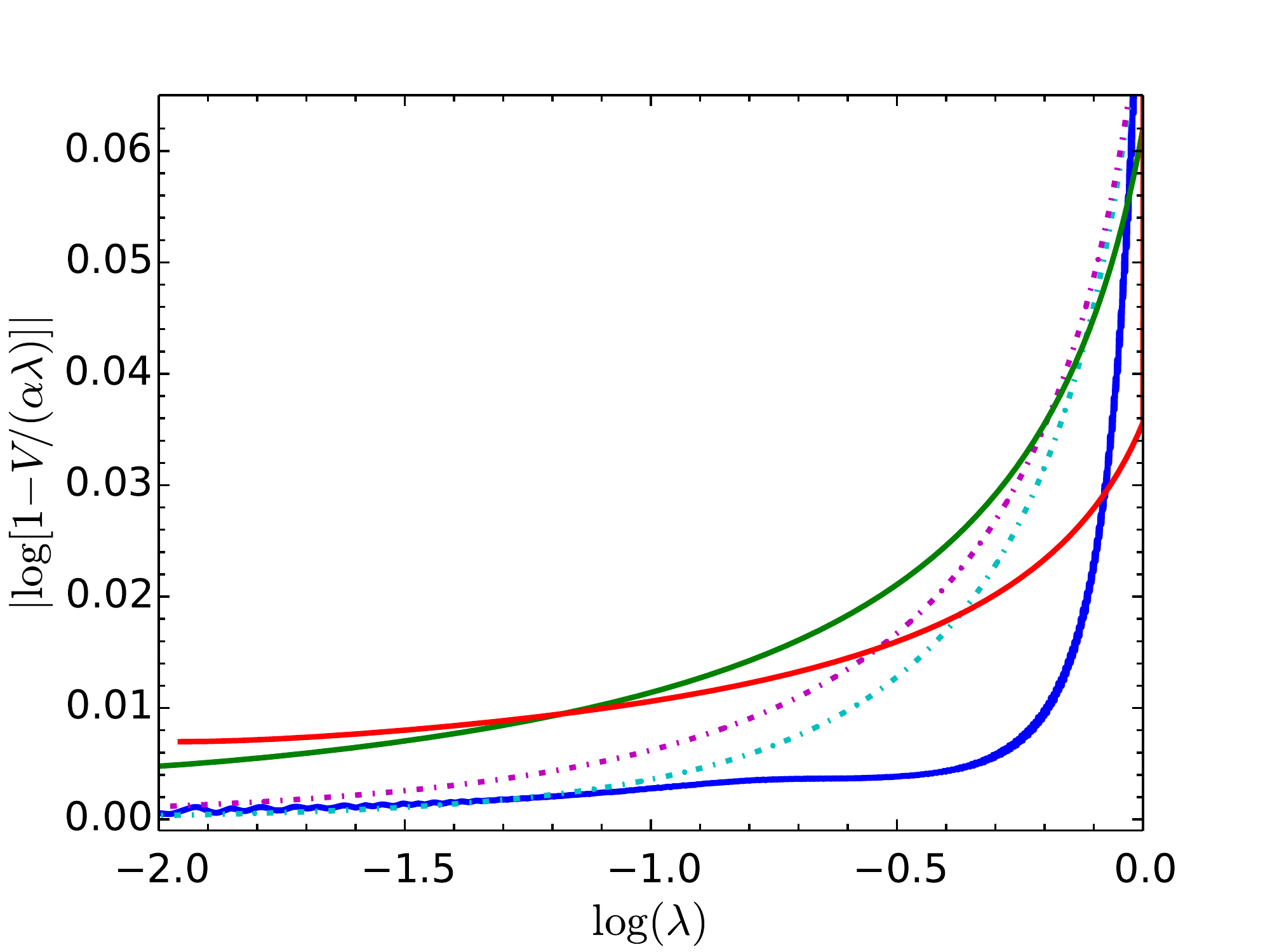} \caption{$|$log[$1-V/(\alpha \lambda)$]$|$ for collapse from realistic initial conditions in a $\Lambda\text{CDM}$ cosmology (blue) and from our analytic calculation of scale-free collapse with $\epsilon=1$ (light blue, dot-dashed), $\epsilon=2/3$ (purple, dot-dashed), $\epsilon=1/3$ (green), and $\epsilon=1/6$ (red).}
\label{fig:2.10}
\end{figure}

\begin{figure}
\includegraphics[width=1\columnwidth]{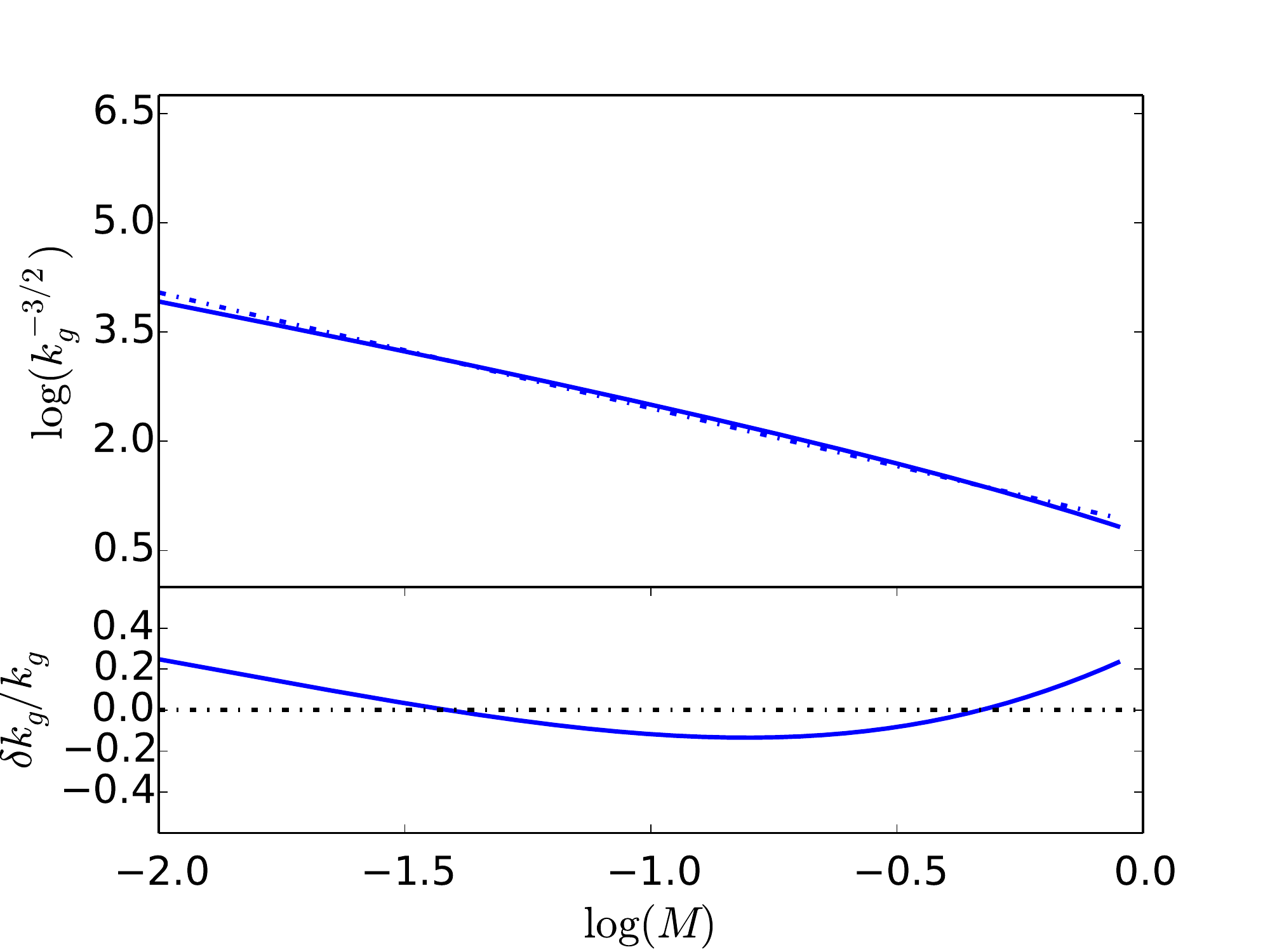}
\includegraphics[width=1\columnwidth]{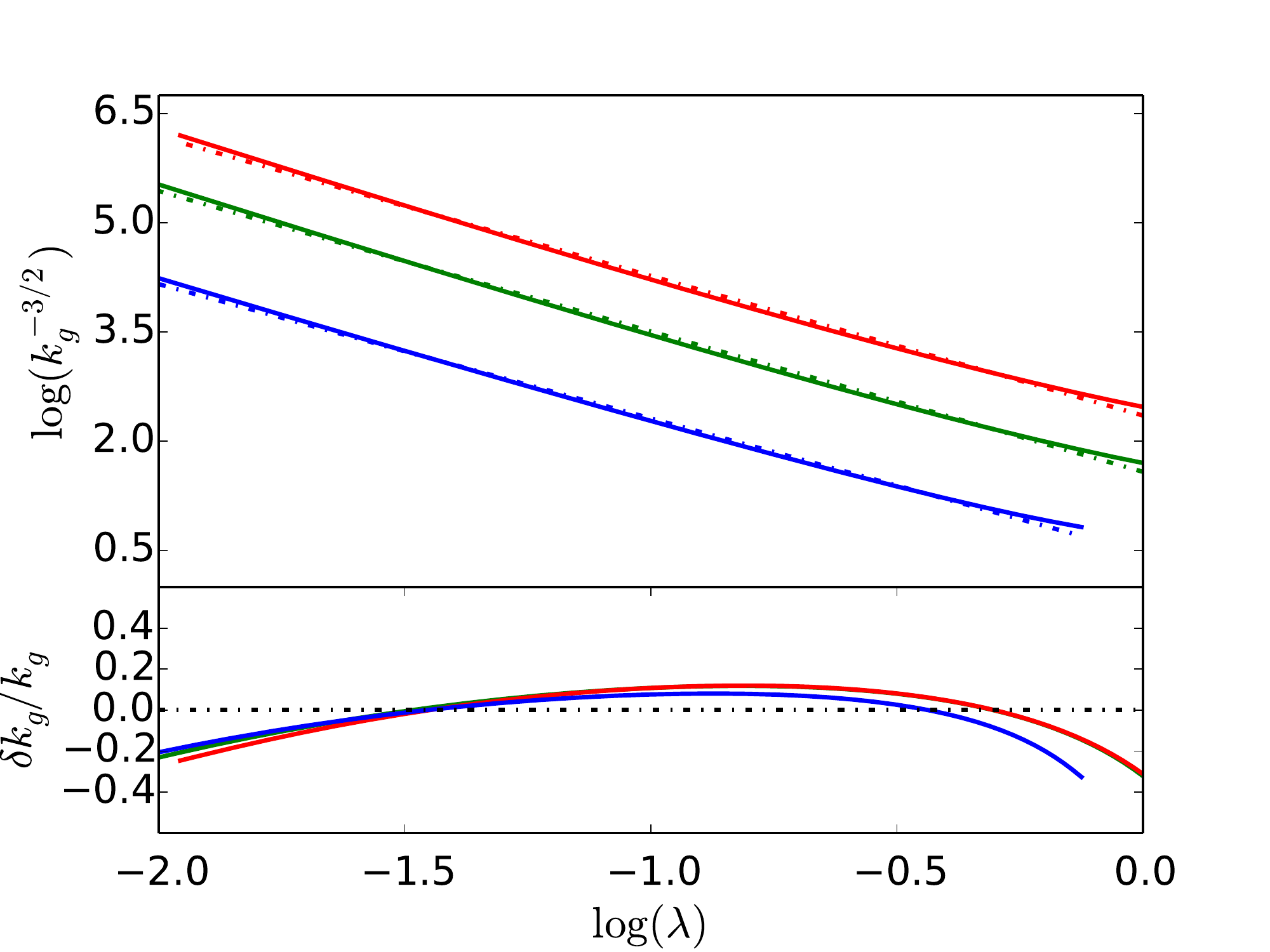}
\caption{Entropy profiles as a function of mass (top) and radius (bottom), along with best-fitting power laws (dot-dashed lines) and residuals, for collapse from realistic initial conditions in a $\Lambda\text{CDM}$ cosmology (blue). Analytic results for scale-free collapse with $\epsilon=1/3$ (green) and $\epsilon=1/6$ (red) are shown in the bottom panel. $\lambda$ is the radius normalised to the fluid shock.}
\label{fig:2.9}
\end{figure}

Interestingly, Figure \ref{fig:2.10} shows that the relative departure from hydrostatic equilibrium is smaller for realistic collapse than for self-similar collapse with $\epsilon=1/3$ or $1/6$. This makes sense in light of the fact that departures from HSE are governed by the accretion rate. The {\it current} accretion rate is low for the realistic case (for $z_{c}=3$, it is more similar to the accretion rate for an $\epsilon=3/5$ perturbation; see Figure \ref{fig:2.6}), so we expect smaller departures from HSE. Hydrostatic equilibrium is therefore a very good approximation deep within realistic halos. This explains, to the extent that $\epsilon\approx \text{const}$ in these inner regions (see Figure \ref{fig:2.6}), why $Q(r)$ is an approximate power law. However, despite the fact that hydrostatic equilibrium is a good approximation deep within the halo, the realistic profile deviates from hydrostatic equilibrium by $\sim 10\%$ at $\text{log}(\lambda)=-0.1$, with the deviation increasing sharply near the shock\footnote{The size of the deviation at $\text{log}(\lambda)=-0.1$ depends weakly on the scheme used to smooth the $\nabla P$ term in the hydrostatic equilibrium equation, but the conclusion that the realistic profile deviates more strongly than the scale-free profiles is unaffected.}. These values are comparable to departures from the equilibrium Jeans equation found in N-body simulations (e.g., \citet{austin05}). 

{\it Cosmology makes little difference.} The upper panel of Figure \ref{fig:2.8} shows that scale-free collapse in a $\Lambda$CDM cosmology results in $k_{g}(M)$ profiles that deviate from power laws in the outer mass shells (which are accreted at low redshifts), as one might expect. However, the magnitude of this deviation is small compared to the variation in the realistic entropy profile. Thus, the departure from self-similarity is dominated by the departure from scale-invariant initial conditions. Indeed, in Figure \ref{fig:2.7}, the power-law index ${\text{d}(\text{log}k_{g})}/{\text{d}(\text{log}\lambda)}$ is very similar in both an EdS and a $\Lambda$CDM cosmology -- the choice of cosmology makes little difference. We can understand this by noting that every overdense shell that eventually collapses behaves like a closed ($\Omega>1$) universe, so $\Omega_{\Lambda}$ contributes a subdominant amount to the total energy of each bound shell. 

{\it Deviations from power-law entropy profiles are similar to N-body results.}  The deviations of our $k_{g}(M)$ and $k_{g}(\lambda)$ profiles from their respective best-fitting power laws are shown in Figure \ref{fig:2.9}; the amplitude of our realistic residual profile is in good agreement with the amplitudes of PPSD power-law residuals found in simulations \citep{navarro10,ludlow10}. Note that, unlike 3D N-body simulations, our idealised fluid model is much less subject to issues of numerical resolution, and it resolves entropy profiles to high precision. Our results imply that the $10$--$20\%$ deviation from a power-law PPSD found in simulations is a real effect. Indeed, plotting the logarithmic slopes produced by our fluid model (Figures \ref{fig:2.8} and \ref{fig:2.7}) as histograms and fitting with Gaussian distributions yields standard deviations of $14.4\%$ for $\text{d}(\text{log}k_{g})/\text{d}(\text{log}M)$ and $12.3\%$ for $\text{d}(\text{log}k_{g})/\text{d}(\text{log}\lambda)$. The characteristic pattern of residuals in Figure \ref{fig:2.9} stems from the progressive steepening (flattening) of the entropy profile with mass (radius) (Figures \ref{fig:2.8} and \ref{fig:2.7}, respectively). Thus, the best-fitting power-law profile for $k_{g}(M)$ is too steep in the outer regions and too shallow in the inner regions, and vice versa for $k_{g}(\lambda)$.

We note that the power-law residuals for $k_{g}(\lambda)$ are very similar for scale-free and realistic initial conditions, even though the departure from a power law is determined entirely by bulk inflow in the scale-free case. For realistic collapse, smaller deviations from HSE (Figure \ref{fig:2.10}) compensate for the lack of scale-invariant initial conditions.

{\it Rough agreement with the Bertschinger power-law index is a red herring.} We regard the fact that the PPSD power-law fits are very close to the \citet{bertschinger85} prediction $Q(r) \propto r^{-1.875}$ (since $Q \propto k_{\rm g}^{-3/2}$, this corresponds to $k_{\rm g} \propto r^{5/4}$) to be purely coincidental and of no real physical significance. The Bertschinger index applies to the secondary infall model ($\epsilon=1$), which is well outside the range of $\epsilon(M)$ for a realistic overdensity profile (see Figure \ref{fig:2.6}). Moreover, the Bertschinger result is an {\it asymptotic} power-law index. As we have seen, convergence to asymptotic slopes is very slow and occurs at small radii that are inaccessible to N-body simulations. The logarithmic entropy slope for realistic collapse has a broad range of values (Figure \ref{fig:2.7}); the Bertschinger value just happens to lie in the middle of this range. 

\begin{figure}
\centering
\includegraphics[width=1\columnwidth]{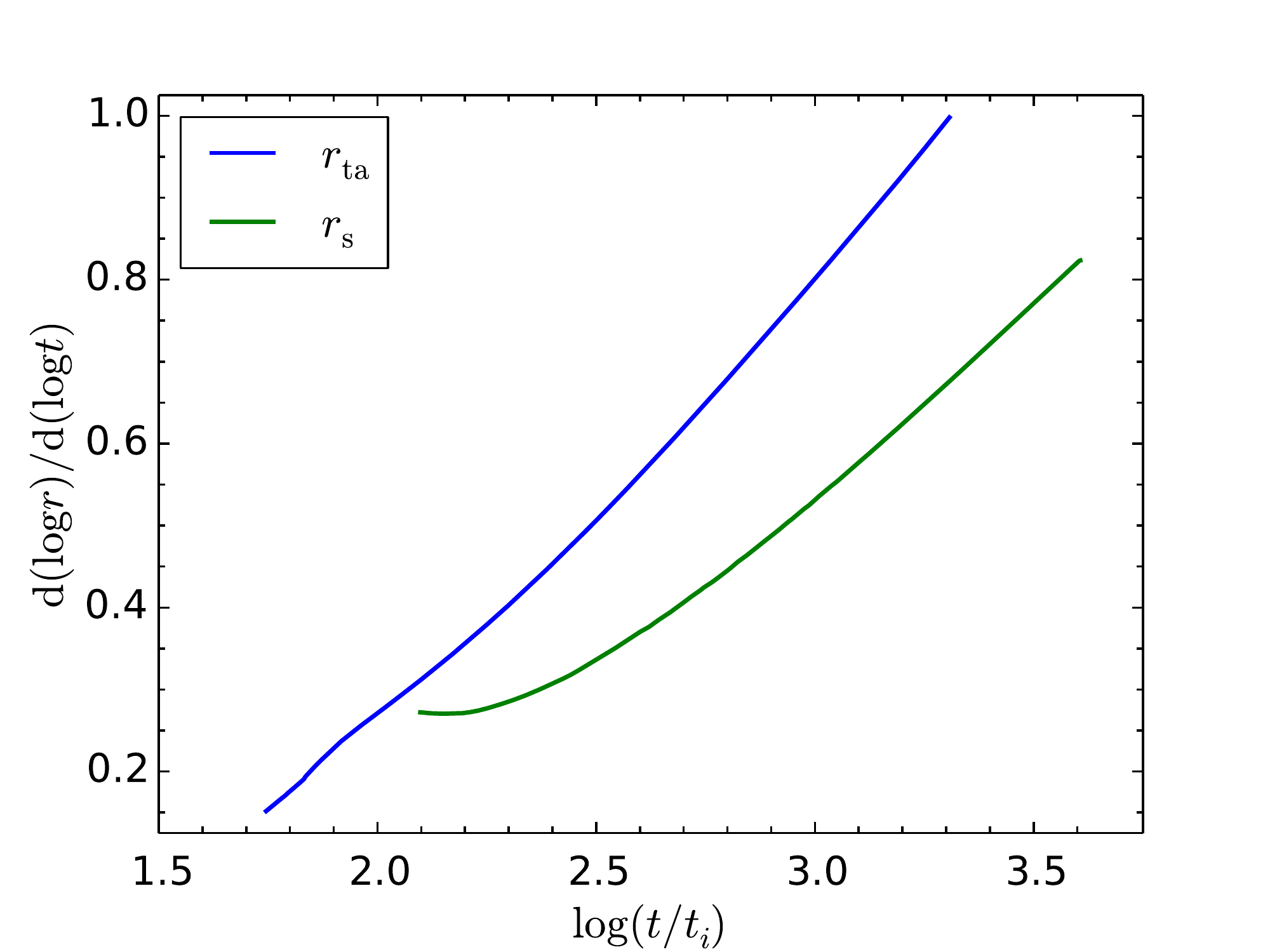} \caption{$\text{d}(\text{log}\rta)/\text{d}(\text{log}t)$ (blue) and $\text{d}(\text{log}\rshock)/\text{d}(\text{log}t)$ (green) from our simulation of halo collapse with realistic initial conditions in a $\Lambda\text{CDM}$ universe. $\rta(t)$ and $\rshock(t)$ are \emph{not} power-law functions of time, which implies that realistic halo collapse does not converge to a self-similar solution.}
\label{fig:2.11}
\end{figure}

{\it Realistic collape is not self-similar.} The realistic case does not have any `hidden' self-similar or scale-free behaviour. Neither $k_{g}(M)$ nor $k_{g}(\lambda)$ are power laws. Moreover, the lack of self-similarity can be seen from the fact that scale radii such as the turnaround radius $r_{\rm ta}(t)$ and the shock radius $r_{\rm s}(t)$ are {\it not} power-law functions of time; Figure \ref{fig:2.11} shows the running power-law behaviour. The power-law scaling $r(t)\propto t^{\alpha}$ is a requirement for self-similar solutions \citep{barenblatt96}, indicating the presence of a conserved quantity $A\equiv r(t)/t^{\alpha}$; the exponent $\alpha$ can either be derived by dimensional analysis (Type I similarity; for example, energy conservation in a Taylor-Sedov explosion) or by solving an eigenvalue problem (Type II similarity; for example, the implosion of a spherical shock wave, as in \citet{guderley42}). The fact that $\rta(t)$ and $\rshock(t)$ are not power laws strengthens the case that realistic collapse does not display any `intermediate asymptotics' \citep{barenblatt96}, which allow convergence to a self-similar solution from initial conditions that are not scale free.


\section{Discussion} 
\label{sect:conclusions} 

Our bottom line is easy to summarise: the pseudo phase space density profile in CDM halos is only a pseudo power law. The $10$--$20\%$ deviation from a power-law PPSD found in N-body simulations is a real effect with physical origins, including deviations from hydrostatic equilibrium within collapsed halos and scale-dependent initial conditions. The fact that the PPSD is intrinsically different from a power law is somewhat obvious based on the results for scale-free fluid collapse: even scale-free initial conditions produce entropy profiles that are not power laws, so we should not expect a power law in the realistic case.

We find that the analogous Lagrangian quantity $k_{g}(M)$ is robust to variations in the velocity profile and instead depends primarily on the initial conditions for halo collapse. Even in a $\Lambda$CDM cosmology, scale-free initial conditions produce $k_{g}(M)$ profiles that are very close to power laws (while they are exact power laws in an EdS cosmology). The realistic $k_{g}(M)$ profile is mainly determined by the function $\epsilon(M) \equiv -{\text{d}(\text{log}[\delta M/M])}/{\text{d}(\text{log}{M})}$ (motivated by $\delta M/M \propto M^{-\epsilon}$ in the self-similar case), which specifies the initial conditions. In particular, there is an intriguing linear relation between $\epsilon(M)$ and the logarithmic slope of the entropy profile: $\text{d}(\text{log}k_{g})/\text{d}(\text{log}M)= \epsilon(M) + 2/3$ (equation \ref{eqn:2.9}). This relation agrees with our realistic simulation to better than $15\%$ over the entire collapsed halo.


An obvious caveat to our conclusions is that we only simulate fluid collapse in 1D, which differs in detail from collisionless collapse in 3D. At the very least, the fluid approximation ignores the effect of velocity anisotropy. It is conceivable that velocity anisotropy somehow conspires to create a true power-law PPSD, rather than a pseudo power law\footnote{For instance, velocity anisotropy clearly affects the density profile, reducing the halo concentration relative to fluid models.}. While we cannot rule out this possibility, we regard it as unlikely. In the absence of a plausible mechanism for accomplishing this result (which has not been proposed), remarkable fine-tuning would be required. Our fluid model gives power-law slopes (Figure \ref{fig:2.4}) and residuals (Figure \ref{fig:2.9}) consistent with those from collisionless N-body CDM simulations \citep{navarro10}. Likewise, we find deviations from hydrostatic equilibrium that are comparable to deviations from the equilibrium Jeans equation found in N-body simulations \citep{austin05}.

The fact that the PPSD is not an exact power law, and that characteristic scales such as the turnaround radius and the shock radius are not power-law functions of time (a fundamental requirement for self-similarity), has several consequences. It implies that the PPSD does not have an elevated status relative to the density or velocity profiles, which show approximate -- but not exact -- universality\footnote{\citet{schmidt08} reach similar conclusions by studying the scatter in PPSD power-law indices for halos on different scales.}. It also implies that we cannot use a strict power-law fit to make detailed dynamical inferences about halo structure, or to extrapolate halo profiles inward, beyond the limits of numerical resolution. Most importantly, since self-similar behaviour is always associated with conserved quantities, the PPSD is unlikely to be a smoking gun indicator of some hitherto unknown integral of motion (such an unknown conserved quantity has been invoked to explain the apparent universality in CDM halo structure, e.g., \citet{pontzen13}). Instead, the PPSD is a rolling power law that runs very slowly and appears approximately linear on a log-log plot because it is a ratio of correlated quantities.

This interpretation is disappointing, but it would not be the first time that an apparent power-law -- thought to be of deep physical significance -- was later found to be a coincidence. One such example is the two-point galaxy correlation function $\xi(r)$. From the 1970s to 2005, galaxy surveys found a power-law $\xi(r)$ that showed no new features in the transition from linear to non-linear scales, or even from non-linear scales to collapsed and virialized groups and clusters. This power law is difficult to achieve in the standard cosmological framework without delicate fine-tuning \citep{berlind02}. SDSS eventually detected a statistically significant deviation from a power law \citep{zehavi04,zehavi05}; today, the power-law correlation function is thought of as a cosmic coincidence, arising in a limited luminosity and redshift range that happened to coincide with the scope of former low-$z$ galaxy surveys \citep{watson11}. We suggest that a similar fate may lie in store for the PPSD.

\section*{Acknowledgments}
We thank Andrey Kravtsov and Matt McQuinn for stimulating conversations. EN and SJ acknowledge a Worster Summer Research Fellowship. We also acknowledge NASA grant NNX15AK81G for support.

\bibliography{master_references}

\end{document}